\documentclass[%
 reprint,
superscriptaddress,  % <<========
 amsmath,amssymb,
 aps,
]{revtex4-1}

\usepackage{graphicx}% Include figure files
\usepackage{dcolumn}% Align table columns on decimal point
\usepackage{bm}% bold math

\begin{document}

\preprint{APS/123-QED}

\title{Ferrimagnetic States of Na-K Alloy Clusters in Zeolite Low-Silica X}% Force line breaks with \\
%\thanks{A footnote to the article title}%

\author{Takehito~Nakano}
\email{takehito.nakano.phys@vc.ibaraki.ac.jp}
\affiliation{Institute of Quantum Beam Science, Graduate School of Science and Engineering, Ibaraki University, \\2-1-1 Bunkyo, Mito, Ibaraki 310-8512, Japan}
\affiliation{Department of Physics, Graduate School of Science, Osaka University, \\1-1 Machikaneyama, Toyonaka, Osaka 560-0043, Japan}
\author{Shingo~Araki}
\email{araki@science.okayama-u.ac.jp}
\affiliation{Department of Physics, Okayama University, Okayama 700-8530, Japan}
\author{Luu~Manh~Kien}
\affiliation{Nano and Energy Center, Hanoi University of Science, Vietnam National University, 334 Nguyen Trai, Thanh Xuan, Hanoi, Vietnam}
\affiliation{Department of Physics, Graduate School of Science, Osaka University, \\1-1 Machikaneyama, Toyonaka, Osaka 560-0043, Japan}
\author{Nguyen~Hoang~Nam}
\affiliation{Center for Materials Science, Faculty of Physics, Hanoi University of Science, Vietnam National University, 334 Nguyen Trai, Thanh Xuan, Hanoi, Vietnam}
\author{Duong~Thi~Hanh}
\affiliation{Department of Physics, Graduate School of Science, Osaka University, \\1-1 Machikaneyama, Toyonaka, Osaka 560-0043, Japan}
\author{Akihiro~Owaki}
\affiliation{Department of Physics, Graduate School of Science, Osaka University, \\1-1 Machikaneyama, Toyonaka, Osaka 560-0043, Japan}
\author{Ken~Goto}
\affiliation{Department of Physics, Graduate School of Science, Osaka University, \\1-1 Machikaneyama, Toyonaka, Osaka 560-0043, Japan}
\author{Akira~Matsuo}
\affiliation{Institute for Solid State Physics, University of Tokyo, \\5-1-5 Kashiwanoha, Kashiwa, Chiba 277-8581, Japan}
\author{Koichi~Kindo}
\affiliation{Institute for Solid State Physics, University of Tokyo, \\5-1-5 Kashiwanoha, Kashiwa, Chiba 277-8581, Japan}
\author{Yasuo~Nozue}
\email{nozue@phys.sci.osaka-u.ac.jp}
\affiliation{Department of Physics, Graduate School of Science, Osaka University, \\1-1 Machikaneyama, Toyonaka, Osaka 560-0043, Japan}

%\affiliation{}
% Authors' institution and/or address\\
% This line break forced with \textbackslash\textbackslash
%}%
%\email{araki@science.okayama-u.ac.jp}
%\author{Second Author}%
% \email{Second.Author@institution.edu}

%\author{Nguyen Hoang Nam}
%\affiliation{%Center for Materials Science, Faculty of Physics, Hanoi University of Science, Vietnam National University,\\334 Nguyen Trai, Thanh Xuan, Hanoi, Vietnam}

%\author{Kota Shimodo}
%\affiliation{%
%Department of Physics, Graduate School of Science, Osaka University, \\1-1 Machikaneyama, Toyonaka, Osaka 560-0043, Japan}

%\collaboration{MUSO Collaboration}%\noaffiliation

%\author{Charlie Author}
% \homepage{http://www.Second.institution.edu/~Charlie.Author}
%\affiliation{
% Second institution and/or address\\
% This line break forced% with \\
%}%
%\affiliation{
% Third institution, the second for Charlie Author
%}%
%\author{Delta Author}
%\affiliation{%
% Authors' institution and/or address\\
% This line break forced with \textbackslash\textbackslash
%}%

%\collaboration{CLEO Collaboration}%\noaffiliation

\date{\today}% It is always \today, today,
             %  but any date may be explicitly specified

\begin{abstract}
In zeolite low-silica X (LSX), $\beta$-cages with the inside diameter of $\approx\,$7 \AA{} are arrayed in a diamond structure. Among them, supercages with the inside diameter of $\approx\,$13 \AA{} are formed and arrayed in a diamond structure by the sharing of windows with the inside diameter of $\approx\,$8 \AA{}. The chemical formula of zeolite LSX used in the present study is given by Na$_{x}$K$_{12-x}$Al$_{12}$Si$_{12}$O$_{48}$ per supercage (or $\beta$-cage), where Na$_{x}$K$_{12-x}$ and Al$_{12}$Si$_{12}$O$_{48}$ are the exchangeable cations of zeolite LSX and the aluminosilicate framework, respectively. Na-K alloy clusters are incorporated in these cages by the loading of guest K metal at $n$K atoms per supercage (or $\beta$-cage).   A N\'eel's N-type ferrimagnetism has been observed at $n = 7.8$ for $x = 4$.  In the present paper, optical, magnetic and electrical properties are studied in detail mainly for $x = 4$. Ferrimagnetic properties are observed at $6.5 < n < 8.5$.  At the same time, the Curie constant suddenly increases.  An optical reflection band of $\beta$-cage clusters at 2.8 eV is observed at $n > 6.5$ in accordance with the sudden increase in the Curie constant.  An electrical resistivity indicates metallic values at  $n \gtrapprox 6$, because a metallic state is realized in the energy band of supercage clusters.  The ferrimagnetism is explained by the antiferromagnetic interaction between the magnetic sublattice of itinerant electron ferromagnetism at supercage clusters and that of localized moments at $\beta$-cage clusters. The electrical resistivity in ferrimagnetic samples at $n = 8.2$ for $x = 4$ increases extraordinarily at very low temperatures, such as $\approx$$10^6$ times larger than the value at higher temperatures.  Observed anomalies in the electrical resistivity resembles the Kondo insulator, but itinerant electrons of narrow energy band of supercage clusters are ferromagnetic differently from the Kondo insulator.
\begin{description}
%\item[Usage]
%Secondary publications and information retrieval purposes.
\item[PACS numbers] 82.75.Vx, 71.28.+d, 75.30.Mb, 75.50.Xx, 75.75.-c, 36.40.-c
%75.75.+a, 75.50.Cc, 75.10.Lp, 73.22.-f, 82.75.Vx
%May be entered using the \verb+\pacs{#1}+ command.
%\item[Structure]
%You may use the \texttt{description} environment to structure your abstract;
%use the optional argument of the \verb+\item+ command to give the category of each item. 
\end{description}
\end{abstract}

%\pacs{Valid PACS appear here}% PACS, the Physics and Astronomy
                             % Classification Scheme.
%\keywords{Suggested keywords}%Use showkeys class option if keyword
                              %display desired
\maketitle

%\tableofcontents

%\section{\label{sec:level1}First-level heading:\protect\\ The line break was forced \lowercase{via} \textbackslash\textbackslash}

%%%%%%%%%%%%%%%%%%%%%%%%%%%%%%%%%%%%%
\section{\label{sec:level1}Introduction}
Zeolite crystals have free spaces of regular cages for guest materials~\cite{Nakano2017-APX}. There are many different types of zeolite structures \cite{IZA}. Alkali metal clusters incorporated in cages of zeolites have a wide variety in electronic properties, such as a ferrimagnetism, a ferromagnetism, an antiferromagnetism, and an insulator-to-metal transition, depending on the kind of alkali metals, their loading density, and the structure type of zeolite frameworks \cite{Nakano2017-APX, Nakano2013ICMInv}. 

In zeolite low-silica X (LSX), supercages and $\beta$-cages with the inside diameters of $\approx\,$13 and $\approx\,$7 \AA{}, respectively,  are arrayed in a diamond structure, namely the double diamond structure.  Up to now, detailed studies have made \cite{Nakano2017-APX, Nakano2013ICMInv, Nakano2006muSR-LSX, Nakano2010-NaLSX, Hanh2010, Nam2010, Nozue2012-Na-LSX, Nakano2013K-LSX, Igarashi2013-Na-LSX, Ikeda2014-Na-LSX, Kien2015, Igarashi2016, Araki2019}.  A N\'eel's N-type ferrimagnetism has been observed in Na-K alloy clusters incorporated into zeolite LSX, where an antiferromagnetic interaction works between nonequivalent magnetic sublattices of supercages and $\beta$-cages \cite{Nakano2017-APX, Hanh2010, Nakano2013ICMInv, Nakano2013K-LSX}.  In the present paper, their optical, electrical and magnetic properties are studied in detail. 

Besides the N\'eel's N-type ferrimagnetism, a ferromagnetism has been observed in Na-rich Na-K alloy clusters in zeolite LSX \cite{Kien2015}.  In pure Na clusters in zeolite LSX, a metallic phase has been observed with the increase in Na loading density \cite{Nakano2017-APX, Nakano2010-NaLSX, Nozue2012-Na-LSX, Igarashi2013-Na-LSX, Igarashi2016}.   In pure K clusters in zeolite LSX, a ferrimagnetic property at higher K loading densities has been observed in a metallic phase \cite{Nakano2017-APX, Nakano2013K-LSX}. Under the pressure loading of K-metal into zeolite LSX, an itinerant electron ferromagnetism has been newly observed at the loading pressure of $\approx\,$0.9 GPa \cite{Araki2019}.  

After the discovery of ferromagnetic properties in K clusters in zeolite A~\cite{Nozue1992-KA},  detailed studies have been made~\cite{Nakano2017-APX, Nakano2013ICMInv, Kodaira1993-KA, Nozue1993-KA, Armstrong1994, Nakano1999-KA, Maniwa1999, Ikeda2000KA, Nakano2000MCLC, Nakano2000, Nakano2001-RbA-KA, Kira2001, Kira2002, Nakano2002ESR, Arita2004, Aoki2004, Ikeda2004KA, Nakano2004, Nakano2007, Nam2007, Nakano2007-HighMag, Nakano2009muSR-KA, Nohara2009, Nohara2011, Nakano-A-2019}.  In zeolite A, $\alpha$-cages with the inside diameter of $\approx\,$11 \AA{} are arrayed in a simple cubic structure. A spin-cant model of Mott-insulator antiferromagnetism of K cluster array in $\alpha$-cages is proposed \cite{Nakano2017-APX, Nakano2004, Nakano2007, Nakano-A-2019}.  In Rb clusters in zeolite A, a ferrimagnetism has been observed~\cite{Nakano2001-RbA-KA, Duan2007, Duan2007b}.   An antiferromagnetism of Mott insulator in alkali metal clusters in sodalite has been clearly observed~\cite{Srdanov1998}, and detailed studies have been made~\cite{Sankey1998, Blake1998, Blake1999, Heinmaa2000, Madsen2001, Tou2001, Scheuermann2002, Madsen2004, Nakamura2009, Nakano2010, Nakano2012-SOD, Nakano2013JKPS, Nakano2013PRB, Nakano2015Mossbauer}.  In sodalite, $\beta$-cages are arrayed in a body centered cubic structure.  Alkali metals in quasi-low-dimensional systems, such as the quasi-one-dimensional metallic system in channel-type zeolite L \cite{Kelly1995, Anderson1997, Thi2016, Thi2017}, has been studied.

%%%%%%%%%%%%%%%%%%%%%%%%%%%%%%%%%%%%%
\subsection{\label{sec:LSX}Zeolite LSX}

\begin{figure}[ht]
\includegraphics[width=7.5cm]{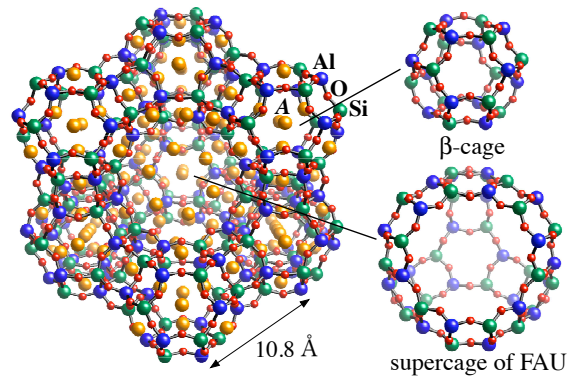}
\caption{\label{fig:FAU}(Color online) Aluminosilicate framework structure of zeolite LSX and typical sites of exchangeable $A$ cations without guest materials. $\beta$-cages are arrayed in a diamond structure. Among them, supercages of FAU are formed and arrayed in a diamond structure.  The distance between adjoining $\beta$-cages (or supercages of FAU) is 10.8 \AA{}. See also the polyhedral illustration of the structure in Fig.~\ref{fig:LSX-poly}.}
\end{figure}

Zeolite X is one of the most typical aluminosilicate zeolites, and is nonmagnetic insulator unless guest materials are loaded.  Zeolite LSX is the zeolite X with Si/Al = 1 in aluminosilicate framework.  The framework of zeolite LSX is negatively charged and illustrated in Fig.~\ref{fig:FAU} together with typical sites of exchangeable monovalent cations ($A$s).  Al and Si atoms are alternately connected by the sharing of O atoms. The space group is $Fd{\bar 3}$ with the lattice constant of 25 \AA{}.  The chemical formula per unit cell is given by $A_{96}$Al$_{96}$Si$_{96}$O$_{384}$ before the loading of guest materials.  The number of cations is the same as that of aluminium atoms in framework.  The framework structure type of zeolite LSX  is called FAU (IUPAC nomenclature \cite{IZA}). The framework of FAU is constructed of $\beta$-cages arrayed in a diamond structure.  Among $\beta$-cages, ``supercages (cavities) of FAU'' are formed and also arrayed in a diamond structure. The distance between adjoining $\beta$-cages (or supercages of FAU) is 10.8 \AA{}. Hereafter, we call ``supercage of FAU'' simply by ``supercage''.  There are eight supercages (or eight $\beta$-cages) in the unit cell, and the chemical formula per supercage (or $\beta$-cage) is given by $A_{12}$Al$_{12}$Si$_{12}$O$_{48}$.  Zeolite LSX used in the present study contains Na and K cations, and the chemical formula per supercage (or $\beta$-cage) is given by Na$_{x}$K$_{12-x}$Al$_{12}$Si$_{12}$O$_{48}$.  Hereafter, we call it by Na$_{x}$K$_{12-x}$-LSX.

\begin{figure}[ht]
\includegraphics[width=5.0cm]{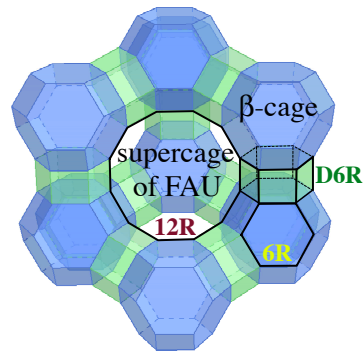}
\caption{\label{fig:LSX-poly} (Color online) Schematic illustrations of framework polyhedra of zeolite LSX. Each $\beta$-cage is connected to four adjoining $\beta$-cages by the sharing of double 6-membered rings (D6Rs), and arrayed in a diamond structure. Supercages of FAU are arrayed in a diamond structure by the sharing of twelve-membered rings (12Rs) with four adjoining supercages.}
\end{figure}

In order to acquire an intuitive understanding of framework structure, a polyhedral form is illustrated in Fig.~\ref{fig:LSX-poly}. Each $\beta$-cage is connected to four adjoining $\beta$-cages by the sharing of hexagonal prisms (double 6-membered rings, D6Rs). Supercages share windows of twelve-membered rings (12Rs) with adjoining supercages.  The inside diameters of 12R and 6R are $\approx\,$8 and $\approx\,$3 \AA{}, respectively.    Each $\beta$-cage shares 6-membered rings (6Rs) with four adjoining supercages.

%%%%%%%%%%%%%%%%%%%%%%%%%%%%%%%%%%%%%
\subsection{\label{sec:level3} Alkali metal loading into zeolite Na$_{x}$K$_{12-x}$-LSX}

Alkali metals are easily loaded into zeolite by the vapor phase for unsaturated condition or by the direct contact with alkali metal for the saturated condition.  In the present paper, we loaded guest K metal at $n$ atoms per supercage (or $\beta$-cage) into Na$_{x}$K$_{12-x}$-LSX, and discribe it as K$_{n}$/Na$_{x}$K$_{12-x}$-LSX. The average number of $s$-electrons provided by the loading of alkali metal is also $n$ per supercage (or $\beta$-cage).  

\begin{figure}[ht]
\includegraphics[width=8.0cm]{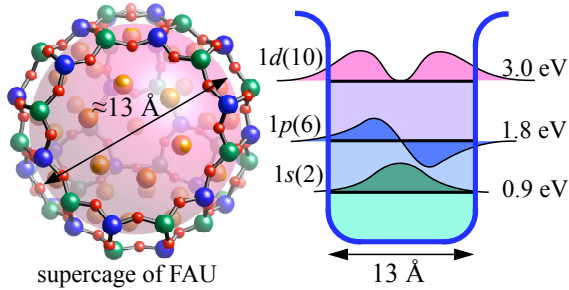}
\caption{\label{fig:cluster1}(Color online) Schematic illustrations of alkali metal cluster in supercage of FAU and the quantum states of $s$-electron in the SQW potential with the diameter of 13 \AA. }
\end{figure}

An outermost $s$-electron of an alkali atom has a large size and a small ionization energy, so that $s$-electrons in bulk alkali metals are well described by the free-electron model.  $s$-electrons introduced in zeolite by the loading of guest alkali atoms move freely over cations distributed in cages.  The aluminosilicate framework, however, is negatively charged and has high-energy conduction bands.  Therefore, $s$-electrons are repulsed by the framework. The $s$-electrons successively occupy quantum states of clusters formed in cages.  If we assume a spherical quantum well (SQW) potential for cage, quantum states, such as 1$s$, 1$p$ and 1$d$ states, are formed in the increasing order of energy, and two, six and ten $s$-electrons can occupy respective quantum states successively \cite{Nakano2017-APX}.  Schematic illustrations of cluster in supercage and quantum states of $s$-electron in the SQW potential with the diameter of 13 \AA{} are given in Fig.~\ref{fig:cluster1}.  A large sphere in supercage is a schematic image of $s$-electron wave function.  1$s$, 1$p$ and 1$d$ quantum states have energies 0.9, 1.8 and 3.0 eV from the bottom of the SQW potential, respectively.  The number in each parentheses indicates the degeneracy including spin.  The optical excitations (dipole transitions) are allowed between 1$s$-and-1$p$ and between 1$p$-and-1$d$ states.  That between 1$s$-and-1$d$ is forbidden.

The SQW potential, however, is primitive for the supercage cluster, because of large 12R windows.  The spheres of $s$-electron wave functions in adjoining supercages largely overlap with each other, because the distance between adjoining supercages is 10.8 \AA{} which is shorter than the inside diameter of supercage $\approx\,$13 \AA{}.  Nevertheless, we use 1$s$, 1$p$ and 1$d$ quantum states of the SQW potential, because of a convenient model to think about quantum states localized in supercage.    In zeolite A, K clusters are well localized in $\alpha$-cages with the inside diameter of $\approx$11 \AA{}, and the SQW model well explains experimental results, because of rather narrow windows of $\alpha$-cages \cite{Nakano2017-APX, Kodaira1993-KA, Nozue1993-KA, Nakano-A-2019}. Electrons in regular supercages of zeolite LSX are expected to construct the energy band, if the contributions of the electron-phonon interaction and the electron correlation are not significant. Because the supercage has the $T_d$ symmetry which has no inversion symmetry at the cage center, 1$s$, 1$p$ and 1$d$ states hybridize with each other.  The electronic states of energy band are constructed of these hybridized states depending on the positions in the Brillouin zone. For example, the electronic states at the bottom of the lowest band are mainly constructed of 1$s$ states.

\begin{figure}[ht]
\includegraphics[width=6.0cm]{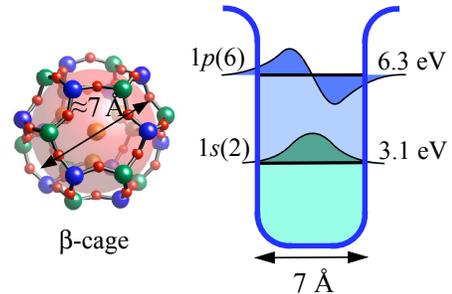}
\caption{\label{fig:cluster2} (Color online) Schematic illustrations of alkali metal cluster in $\beta$-cage and the quantum states of $s$-electron in the SQW potential with the diameter of 7 \AA{}.}
\end{figure}

Schematic illustrations of cluster in $\beta$-cage and quantum states of $s$-electron in the SQW potential with the diameter of 7 \AA{} are given in Fig.~\ref{fig:cluster2}.  A large sphere in $\beta$-cage is a schematic image of $s$-electron wave function.  1$s$ and 1$p$ quantum states have energies 3.1 and 6.3 eV from the bottom of the SQW potential, respectively. These energies are much higher than respective states in supercage, because of a narrow size of $\beta$-cage.  As adjoining $\beta$-cages are well separated by D6Rs as shown in Fig.~\ref{fig:LSX-poly}, $s$-electron wave functions in adjoining $\beta$-cages scarcely overlap with each other, but a finite overlap occurs through 6Rs between supercages and $\beta$-cages.

%%%%%%%%%%%%%%%%%%%%%%%%%%%%%%%%%%%%%
\subsection{\label{sec:level4} Electronic properties of Na-K alloy clusters in K$_{n}$/Na$_{x}$K$_{12-x}$-LSX}

Electronic properties of Na-K alloy clusters in K$_{n}$/Na$_{x}$K$_{12-x}$-LSX largely depend on $x$ as well as $n$.  The contributions of Na atoms are the larger ionization energy and the smaller cation size, compared with those of K atoms.  In K$_{n}$/K$_{12}$-LSX (namely $x = 0$), pure K clusters show a metallic phase at $n \gtrapprox 6$ and a ferrimagnetic property at the saturation loading density $n \approx 9$ at ambient pressure \cite{Nakano2017-APX, Nakano2013K-LSX}. Under the pressure loading of K-metal into zeolite K$_{12}$-LSX, the disappearance of the ferrimagnetism has occurred and an itinerant electron ferromagnetism have been newly observed at $n \approx 15$ at the loading pressure $\approx\,$0.9 GPa \cite{Araki2019}. In Na-K alloy clusters in K$_{n}$/Na$_{4}$K$_{8}$-LSX (namely $x = 4$), the N\'eel's N-type ferrimagnetism has been observed \cite{Nakano2017-APX, Nakano2013ICMInv, Hanh2010, Nakano2006muSR-LSX}. Under the pressure loading of K-metal into zeolite Na$_{4}$K$_{8}$-LSX, a new ferrimagnetism have been observed at the loading pressure $\approx\,$0.5 GPa \cite{Nam2010}. In K$_{n}$/Na$_{7.3}$K$_{4.7}$-LSX (namely $x = 7.3$), a nearly pure ferromagnetism in an insulating phase has been observed at $n \approx 9$ \cite{Kien2015}.  The origin of the ferromagnetism is assigned to the ferromagnetic superexchange coupling between magnetic moments at $\beta$-cage clusters via $sp^3$ closed-shell clusters at supercages. 

Pure Na clusters are generated by the Na metal loading into zeolite Na$_{12}$-LSX (namely $x = 12$). Insulating and non-magnetic states of pure Na clusters have been observed in Na$_{n}$/Na$_{12}$-LSX for $n \lessapprox 11$. A metallic phase has been observed with the increase in $n$.  A thermally activated paramagnetic susceptibility has been observed significantly at $n \approx 16$, and is assigned to the thermal distribution of metastable small polarons \cite{Nakano2017-APX, Nakano2010-NaLSX, Nozue2012-Na-LSX}. The temperature dependence of the paramagnetic susceptibility has been observed in the shift of $^{23}$Na NMR narrow line \cite{Igarashi2013-Na-LSX, Igarashi2016}, although there are many nonequivalent Na sites in Na$_{n}$/Na$_{12}$-LSX \cite{Nakano2017-APX, Ikeda2014-Na-LSX}. This result indicates that Na cations are hopping thermally over many Na sites at higher temperatures during the NMR time window, and nuclei of relevant Na cations feel average paramagnetic field of thermally metastable small polarons. 

In the present paper, optical, magnetic and electrical properties in K$_{n}$/Na$_{x}$K$_{12-x}$-LSX are studied in detail mainly for $x = 4$.  Ferrimagnetic properties are observed at $6.5 < n < 8.5$ in K$_n$/Na$_4$K$_{8}$-LSX. At the same time, the Curie constant suddenly increases, and a reflection band of $\beta$-cage clusters at 2.8 eV is observed at $n > 6.5$.  An electrical resistivity indicates metallic value at  $n \gtrapprox 6$.  The electrical resistivity increases extraordinarily at very low temperatures in ferrimagnetic samples, such as $\approx\,$$10^6$ times larger than the value at higher temperatures. The ferrimagnetism is explained by the antiferromagnetic interaction between the magnetic sublattice of itinerant electron ferromagnetism at supercage clusters and that of localized moments at $\beta$-cage clusters.  We try to explain these anomalies of electrical resistivity by the analogy of the Kondo insulator, where itinerant electron spins of supercage clusters interact with localized electron spins of $\beta$-cage clusters.  Itinerant electrons of narrow energy band of supercage clusters, however, is ferromagnetic, differently from the Kondo insulator.

%%%%%%%%%%%%%%%%%%%%%%%%%%%%%%%%%%

\section{Experimental Procedures}

Zeolites are crystalline powder of few microns in grain size.  The as-synthesized zeolite LSX was $x = 9$.  Na cations were fully exchanged to K$_{12}$-LSX in KCl aqueous solution.  K$_{12}$-LSX was partly ion-exchanged in aqueous NaCl solution in order to get Na$_{x}$K$_{12-x}$-LSX. The value of $x$ was estimated by means of inductively coupled plasma (ICP) spectroscopy.  Zeolite Na$_{x}$K$_{12-x}$-LSX was fully dehydrated in vacuum at 500$^\circ$C for one day.  Distilled potassium metal was set into a quartz glass tube together with the dehydrated Na$_{x}$K$_{12-x}$-LSX in a glovebox filled with a pure He gas containing less than 1 ppm of O$_2$ and H$_2$O.  The potassium metal in the quartz glass tube was adsorbed into the Na$_{x}$K$_{12-x}$-LSX at 150$^\circ$C. The thermal annealing was made for enough time to get the homogeneous K-loading.  The value of $n$ was estimated from the weight ratio of K-metal to Na$_{x}$K$_{12-x}$-LSX powder.

The optical diffuse reflectivity $r$ was measured at room temperature by the use of an FTIR spectrometer (Nicolet Magna 550) and a double monochromator-type UV-vis-NIR spectrometer (Varian Cary 5G). KBr powder was used for the reference of white powder. Since samples are extremely air-sensitive, optical measurements were performed on samples sealed in quartz glass tubes.  The diffuse reflectivity $r$ was transformed to the optical absorption spectrum by the Kubelka-Munk function $(1-r)^2/2r$ which gives the ratio of the absorption coefficient to the reciprocal of powder size.  The sum of the normal reflectivity $R$ and the transmission coefficient $T_{\rm{r}}$ was obtained by the transformation $R+T_{\rm{r}} = 4r/(1+r)^2$ \cite{Kodaira1993-KA}. The normal reflectivity spectrum was obtained as $R=4r/(1+r)^2$ at the spectral region for $T_{\rm{r}} \ll R$. 

A SQUID magnetometer (MPMS-XL, Quantum Design) was used for magnetic measurements in the temperature range 1.8-300 K. A diamagnetic signal from the quartz glass tube is included in the SQUID signal as the temperature-independent background, and is subtracted from measured magnetization. 

For an electrical resistivity measurement, powder samples were put between two gold electrodes, and an adequate compression force $\approx\,$1 MPa was applied during the measurements.  Because of the extreme air-sensitivity of samples, they were kept in a handmade air-proof cell. These setting procedures were completed inside the glovebox. The cell was set into Physical Property Measurement System (PPMS, Quantum Design), and the temperature was changed between 2 and 300 K. The electrical resistivity of the cell was measured by the four-terminal method with the use of Agilent E4980A LCR meter at the frequency range from 20 Hz to 2 MHz and DC. The frequency dependence of the complex impedance was analyzed by the Cole-Cole plot, and the DC or 20 Hz electrical resistivity $\rho$ was obtained by the multiplication of the dimensional factor (area/thickness) of compressed powder. Due to the constriction resistance \cite{Holm1967} at connections between powder particles as well as the low filling density of powder particles, the observed resistivity is about two orders of magnitude larger than the true value. The relative values in different samples, however, can be compared with each other within an ambiguity of factor, because of the constant compression force.  Fortunately, values in the present study change in the several orders of magnitude. Detailed experimental procedures are explained elsewhere \cite{Nozue2012-Na-LSX}.  The upper limit of the present resistivity measurement was $\approx\,$10$^9$ $\Omega\,$cm, and obtained values for $\rho \gtrapprox$ 10$^9$ $\Omega\,$cm are unreliable. The ionic conductance of dehydrated zeolites under the low compression force is expected in the order of 10$^{-9}$ $\Omega^{-1}\,$cm$^{-1}$ at room temperature \cite{Kelemen1992},  and is negligible at lower temperatures in the present study. A small resistivity of the short circuit in the cell  ($<$ 0.1 $\Omega\,$cm) is included in the measured value, but is negligible in the present study.

The high-field magnetization was measured by using an induction method with a multilayer pulse magnet at the Institute for Solid State Physics, the University of Tokyo.  A non-destructive pulsed magnet for 70 T was used for this measurement. Sample sealed in a high-quality quartz glass tube with a diameter of 2 mm was set in the pickup coils.  The observed magnetization is normalized by the results obtained by the SQUID magnetometer at $H < 5 \times 10^4$ Oe. 

%%%%%%%%%%%%%%%%%%%%%%%%%%%%%%%%%%%

\section{Experimental Results}

\subsection{\label{sec:Optical}Optical properties}

\begin{figure}[ht]
\includegraphics[width=6.5cm]{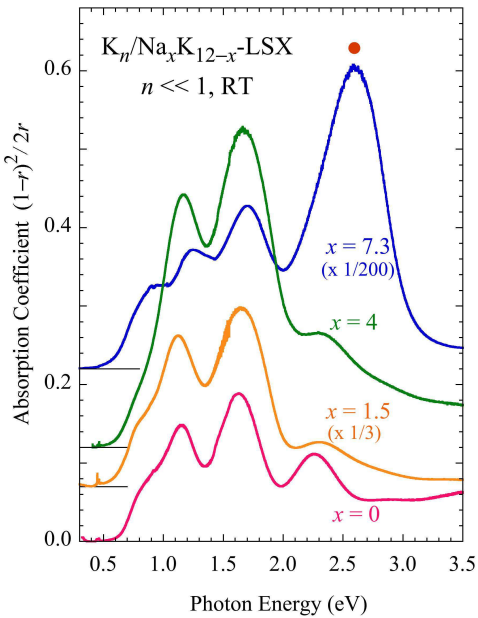}
\caption{\label{fig:Abs}(Color online) Absorption spectra of dilutely K-loaded K$_n$/K$_{12}$-LSX ($x = 0$), K$_n$/Na$_{1.5}$K$_{10.5}$-LSX ($x = 1.5$), K$_n$/Na$_4$K$_{8}$-LSX ($x = 4$) and K$_n$/Na$_{7.3}$K$_{4.7}$-LSX ($x = 7.3$) at room temperature, where $n \ll 1$.  }
\end{figure}

Optical resonant absorption and reflection spectra provide an important information on the dipole transition of electronic states including nonmagnetic ones.  Absorption spectra of dilutely K-loaded K$_{n}$/Na$_{x}$K$_{12-x}$-LSX ($n \ll 1$) at room temperature (RT) are shown in Fig.~\ref{fig:Abs} for  $x = 0$, 1.5, 4 and 7.3. Spectra in K$_n$/K$_{12}$-LSX, K$_n$/Na$_{1.5}$K$_{10.5}$-LSX and K$_n$/Na$_4$K$_{8}$-LSX have continuous peaks above $\approx\,$0.6 eV. These peaks are assigned to the excitation from 1$s$-like states to the empty energy bands of supercage network \cite{Nakano2013ICMInv}.   A new band appears at $\approx\,$2.6 eV with mark in K$_n$/Na$_{7.3}$K$_{4.7}$-LSX, in addition to above mentioned continuous peaks.  This new band is assigned to the excitation from 1$s$-like states to 1$p$-like ones of clusters in $\beta$-cages \cite{Kien2015}.  

If we assume strict SQW potentials shown in Figs.~\ref{fig:cluster1} and \ref{fig:cluster2}, the 1$s$--1$p$ excitation energies are expected at 0.9 and 3.2 eV in clusters localized in supercage and $\beta$-cage, respectively. Because of the lack of the inversion symmetry at the center of supercage, 1$s$, 1$p$ and 1$d$ states hybridize partly with each other in the energy band.  Continuous DOS of the hybridized energy band of supercage network are expected, because of the electron transfer through large 12R windows with the size $\approx\,$$8$ \AA{}.   In principle, the absorption coefficient of the band-to-band excitation is proportional to the joint-density-of-states times the transition dipole moments between ground states and excited states. The observed gap energy of continuous absorption bands, $\approx\,$0.6 eV, originates from the formation energy of small bipolarons at supercages at low K-loading densities, as stated in Section~\ref{sec:tUSn}. Small bipolarons are optically excited to the extended states of the hybridized energy band of supercage network. The $\beta$-cage potential provides well-isolated electronic states, because of narrow windows. The optical excitation from 1$s$ to 1$p$ states is expected at 3.2 eV in Fig.~\ref{fig:cluster2}, but the effective potential size is expected to be slightly larger than 7 \AA{}, such as 7.8 \AA{}, in order to fit the observed excitation energy $\approx\,$2.6 eV.  As discussed in Section~\ref{sec:Cluster-beta}, the surrounding cations are expected to extend the confinement potential.

\begin{figure}[ht]
\includegraphics[width=6.5cm]{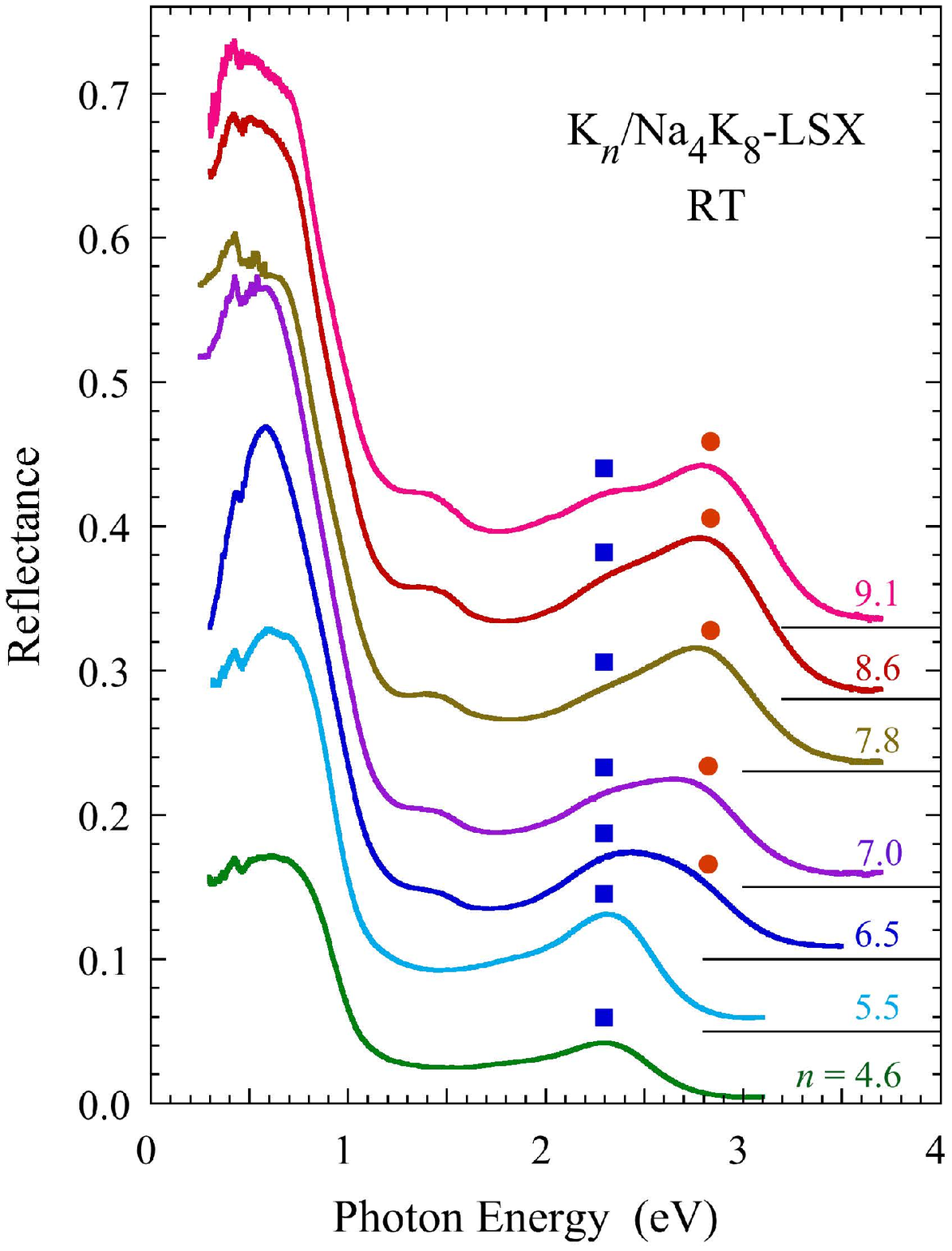}
\caption{\label{fig:Ref-4}(Color online) Reflection spectra of K$_n$/Na$_4$K$_{8}$-LSX at room temperature. The value of $n$ is indicated for each spectrum.}
\end{figure}

Reflection spectra of K$_n$/Na$_4$K$_{8}$-LSX ($x = 4$) at room temperature are shown in Fig.~\ref{fig:Ref-4}. The K-loading density $n$ is indicated for each spectrum.  A reflection band of nearly metallic $s$-electrons of supercage clusters is seen below $\approx$1 eV in each spectrum. The plasma edge of metallic $s$-electrons is estimated to be $\approx$1 eV. With the increase in $n$, the $\beta$-cage cluster bands grow around $\approx$2.3 and $\approx$2.8 eV.  The 2.3 eV band grows at lower values of $n$. As shown in Section~\ref{sec:Magnetic}, a ferrimagnetism and a sudden increase in the Curie constant are observed simultaneously at $6.5 < n < 8.5$.  The 2.8 eV band of $\beta$-cage clusters is assigned to the magnetic K-rich clusters (small polarons) for $6.5 \lessapprox n \lessapprox 8.5$ and nonmagnetic K-rich clusters (small bipolarons) for $8.5 \lessapprox n$, as discussed in Section~\ref{sec:Cluster-beta}. The 2.3 eV band is assigned to nonmagnetic Na-rich clusters at $\beta$-cages. 

\begin{figure}[ht]
\includegraphics[width=6.5cm]{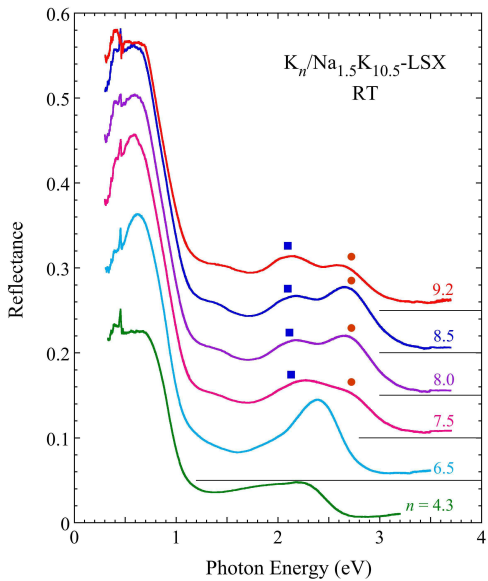}
\caption{\label{fig:Ref-1.5}(Color online) Reflection spectra of K$_n$/Na$_{1.5}$K$_{10.5}$-LSX at room temperature. The value of $n$ is indicated for each spectrum.}
\end{figure}

In K$_n$/Na$_{1.5}$K$_{10.5}$-LSX ($x = 1.5$), similar reflection spectra are observed at room temperature, as shown in Fig.~\ref{fig:Ref-1.5}.  Reflection bands of $\beta$-cage clusters are observed at similar energies 2.2 and 2.8 eV. The 2.8 eV band appears at $n \gtrapprox 7.5$.  As shown in Section~\ref{sec:Magnetic}, a ferrimagnetism and an increase in the Curie constant are observed simultaneously at $7.8 < n \lessapprox 9.5$.  The 2.8 eV band is assigned to the K-rich magnetic clusters (small polarons) at $\beta$-cages, as discussed in Section~\ref{sec:Cluster-beta}.   Reflection bands at 2.2, 2.3 and 2.4 eV are expected to be nonmagnetic Na-rich $\beta$-cage clusters with different configurations of cations.

%%%%%%%%%%%%%%%%%%%%%%%%%%%%%%%%%%%
\subsection{\label{sec:Magnetic}Magnetic properties}

\begin{figure}[ht]
\includegraphics[width=6.5cm]{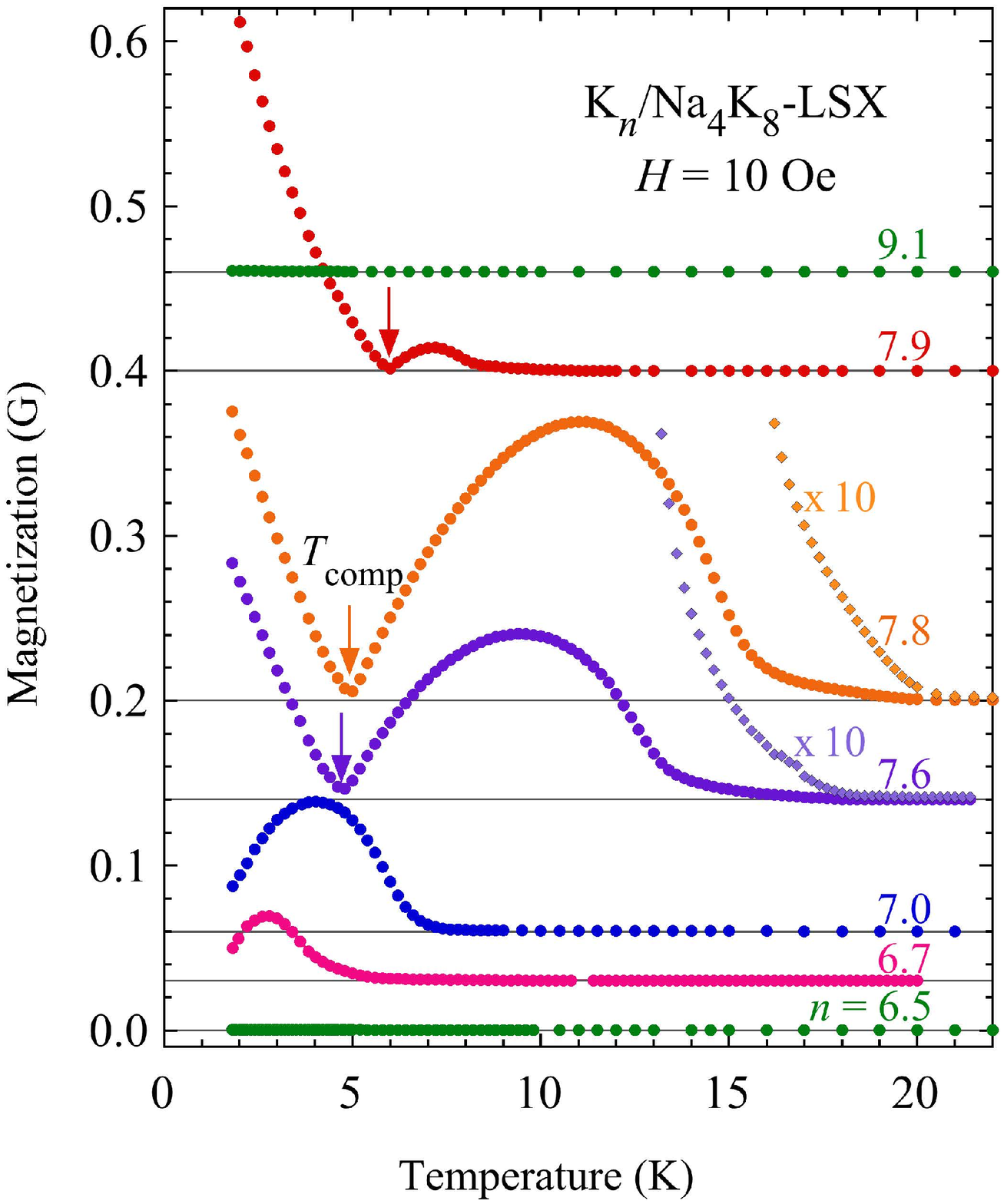}
\caption{\label{fig:MT_x=4}(Color online) Temperature dependences of magnetization in K$_n$/Na$_4$K$_{8}$-LSX under the magnetic field of 10 Oe. The value of $n$ is indicated for each curve.}
\end{figure}

Temperature dependences of magnetization in K$_n$/Na$_4$K$_{8}$-LSX under the low magnetic field of 10 Oe are shown in Fig.~\ref{fig:MT_x=4}. The value of $n$ is indicated for each curve. The observed large magnetization originates from the spontaneous magnetization, because of an applied magnetic field is very weak.  The Curie temperature increases and decreases with $n$. A typical N\'eel's N-type ferrimagnetism with the zero minimum of magnetization at the compensation temperature $T_{\rm{comp}}$ is seen at $n = 7.6$, 7.8 and 7.9. A similar zero minimum may be expected below 1.8 K at $n = 6.7$ and 7.0.  A gradual increase in magnetization around the Curie temperature is seen at $n = 7.6$ and 7.8 with the decrease in temperature, indicating that a weak inhomogeneity is expected to exist in the temperature of the magnetic phase transition.  The zero minimum at $T_{\rm{comp}}$, however, is clearly seen.  

The N\'eel's N-type ferrimagnetism is explained by an antiferromagnetic interaction between two nonequivalent magnetic sublattices A and B, one of which (A) has both a very weak internal magnetic interaction and the saturation magnetization which is larger than the magnetization of the other sublattice (B). The sublattice B has a stronger internal interaction. Below the Curie temperature, the sublattice B increases the spontaneous magnetization. The magnetization of sublattice A follows the sublattice B with the opposite direction.  At $T_{\rm{comp}}$, magnetizations of sublattices A and B have the same magnitude with opposite directions, and the total magnetization becomes zero. Below $T_{\rm{comp}}$, the sublattice A has the magnetization larger than that of sublattices B.  As discussed later in Section~\ref{sec:SublatticeInteraction}, we introduce a model of two magnetic sublattices A and B constructed by localized magnetic moments of $\beta$-cage clusters and an itinerant electron ferromagnetism of supercage clusters, respectively.  In Fig.~\ref{fig:MT_x=4}, $T_{\rm{comp}}$ seems to approach the Curie temperature relatively, indicating that an antiferromagnetic interaction between magnetic sublattices A and B and/or the magnetization of sublattice A increase with $n$ at the ferrimagnetic condition.

\begin{figure}[ht]
\includegraphics[width=7.5cm]{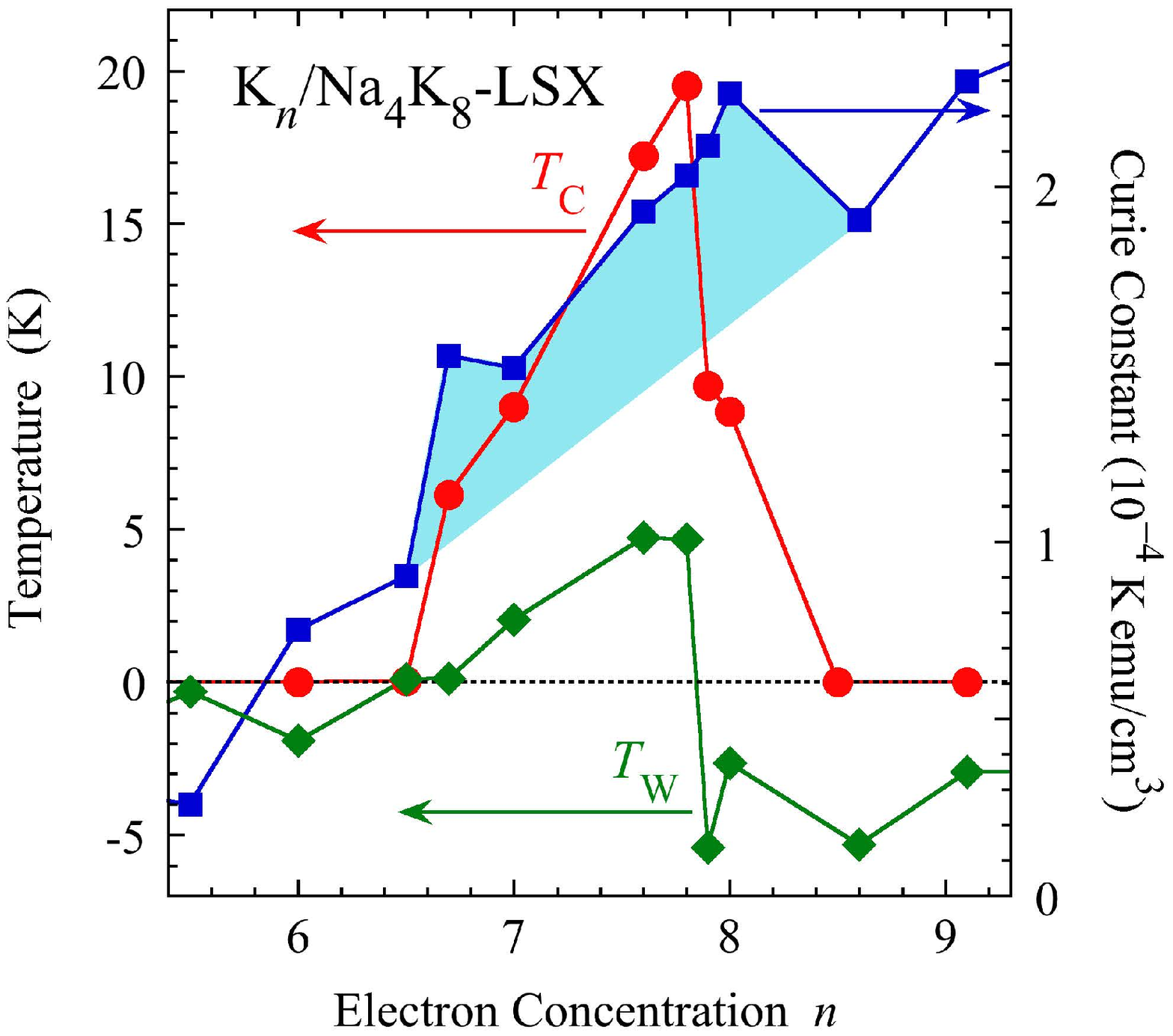}
\caption{\label{fig:x=4TcC}(Color online)  $n$-dependences of the asymptotic Curie temperature $T_{\rm{C}}$, the Weiss temperature $T_{\rm{W}}$ and the Curie constant in K$_n$/Na$_4$K$_{8}$-LSX.}
\end{figure}

$n$-dependences of the asymptotic Curie temperature $T_{\rm{C}}$, the Weiss temperature $T_{\rm{W}}$ and the Curie constant in K$_n$/Na$_4$K$_{8}$-LSX are shown in Fig.~\ref{fig:x=4TcC}. The Curie constant has a sudden increase at the ferrimagnetic condition $6.5 < n < 8.5$, as colored in blue.  $T_{\rm{W}}$ is positive and negative at lower and higher values of $n$, respectively.  The 2.8 eV band of $\beta$-cage clusters grows at $n \gtrapprox 6.5$ in Fig.~\ref{fig:Ref-4} in accordance with the sudden increase in the Curie constant.  
 
The sudden increase of the Curie constant in Fig.~\ref{fig:x=4TcC} is estimated to be $\approx\,$$5 \times 10^{-5}$ K$\,$emu/cm$^3$.  If we assume localized magnetic moments of $\beta$-cage clusters with spin $s = 1/2$ and $g = 2$, the Curie constant $C_\beta$ is given by
\begin{eqnarray} 
{C_\beta } = \frac{{{N_\beta }{g^2}\mu _{\rm{B}}^2s(s + 1)}}{{3{k_{\rm{B}}}}}= \frac{{{N_\beta }\mu _{\rm{B}}^2}}{{{k_{\rm{B}}}}},
\label{CurieConstBeta}
\end{eqnarray}
where $N_\beta$ and $k_{\rm{B}}$ are the number density of magnetic clusters at  $\beta$-cages and the Boltzmann constant, respectively.   The estimated value of $N_\beta$ amounts to $\approx$15\% of $\beta$-cages and the saturation magnetization becomes $\approx\,$0.7 G.   

\begin{figure}[ht]
\includegraphics[width=7.5cm]{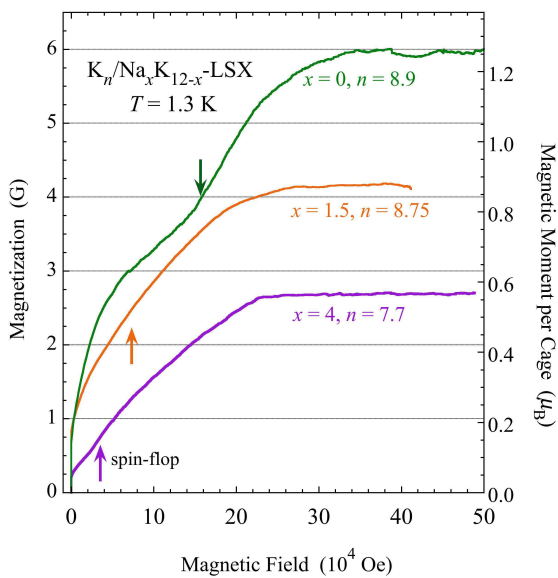}
\caption{\label{fig:MH-highH}(Color online) The magnetization process up to high magnetic fields at 1.3 K for K$_n$/Na$_{x}$K$_{12-x}$-LSX, where the respective values of ($x$, $n$) are (4, 7.7), (1.5, 8.75) and (0, 8.9).  The corresponding magnetic moment per supercage (or $\beta$-cage) is indicated in the axis on the right in units of $\mu_B$.}
\end{figure}

The background Curie constant in Fig.~\ref{fig:x=4TcC} is $n$ dependent, for example, $\approx\,$$1.3 \times 10^{-4}$ K$\,$emu/cm$^3$  at $n \approx 7.5$.  The Curie constant of an itinerant electron ferromagnetism for supercage clusters, $C_{\rm{s}}$, is given by
\begin{eqnarray} 
{C_{\rm{s}}} = \frac{{{N_0}{p_{{\rm{eff}}}}^2{\mu _{\rm{B}}}^2}}{{3{k_{\rm{B}}}}},
\label{CurieConstSuper}
\end{eqnarray}
where $N_0$ and $p_{{\rm{eff}}}\mu _{\rm{B}}$ are the number density of supercages and the effective local magnetic moment per supercage, respectively.  The value of $p_{{\rm{eff}}}$ estimated from the background Curie constant is $\approx\,$1.1 which corresponds to the saturation magnetization of $\approx\,$5.3 G.  In case of the itinerant electron ferromagnetism, however, the spontaneous magnetization at low magnetic fields is much smaller than that estimated from the Curie constant, such as $\approx\,$1/3 in the itinerant electron ferromagnetism in the pressure loading of K metal into K$_{12}$-LSX \cite{Araki2019}. If we assume a similar ratio, the spontaneous magnetization of supercage clusters will be $\approx\,$1.8 G at low temperatures. The total magnetization will be $\approx\,$2.5 G.  At very high magnetic fields, the saturation of total magnetization is observed at 2.7 G as shown later in Fig.~\ref{fig:MH-highH}.  In order to explain the N\'eel's N-type ferrimagnetism observed in Fig.~\ref{fig:MT_x=4}, the spontaneous magnetization of supercage clusters at low temperatures will be smaller than $\approx\,$0.7 G of the saturation magnetization at $\beta$-cage clusters.

\begin{figure}[ht]
\includegraphics[width=6.5cm]{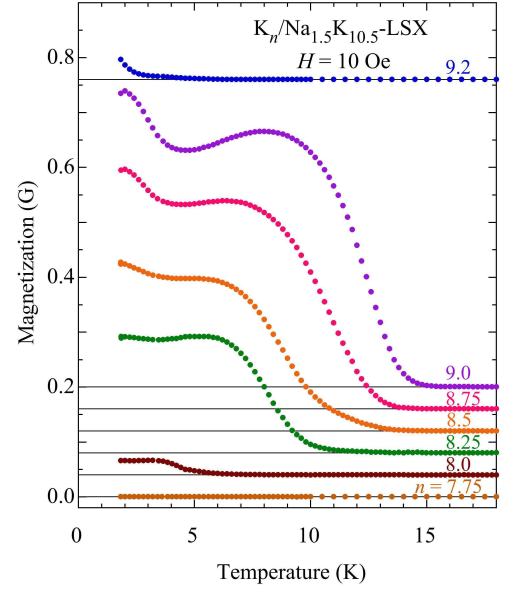}
\caption{\label{fig:MT_x=1.5}(Color online) Temperature dependences of magnetization in K$_n$/Na$_{1.5}$K$_{10.5}$-LSX under the magnetic field of 10 Oe. The value of $n$ is indicated for each curve.}
\end{figure}

\begin{figure}[ht]
\includegraphics[width=7.5cm]{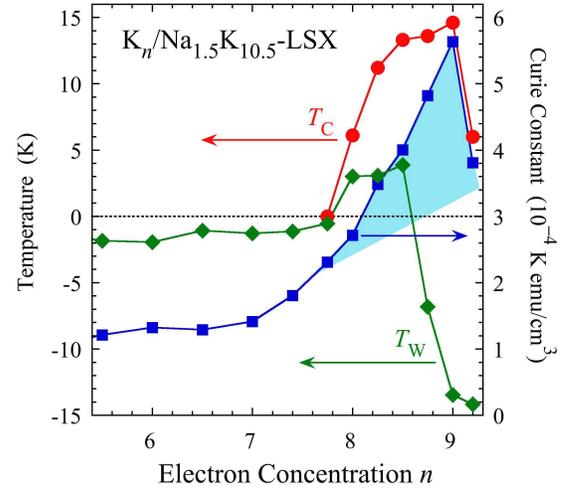}
\caption{\label{fig:x=1.5TcC}(Color online) $n$-dependences of the asymptotic Curie temperature $T_{\rm{C}}$, the Weiss temperature $T_{\rm{W}}$ and the Curie constant in K$_n$/Na$_{1.5}$K$_{10.5}$-LSX.}
\end{figure}

The temperature dependence of magnetization in K$_n$/Na$_{1.5}$K$_{10.5}$-LSX under the magnetic field of 10 Oe is shown in Fig.~\ref{fig:MT_x=1.5}. The value of $n$ is indicated for each curve.  The Curie temperature increases and decreases with $n$.  The magnetization has a minimum at the temperatures lower than the respective Curie temperatures, indicating that this is the N\'eel's P-type ferrimagnetism, where the magnetization of $\beta$-cage clusters is smaller than that of supercage clusters at any temperature.   $n$-dependences of the asymptotic Curie temperature $T_{\rm{C}}$, the Weiss temperature $T_{\rm{W}}$ and the Curie constant are shown in Fig.~\ref{fig:x=1.5TcC}. The Curie constant is much larger than that in K$_n$/Na$_4$K$_{8}$-LSX. The Curie constant has an increase at the ferrimagnetic condition $7.8 < n \lessapprox 9.5$, as colored in blue. $T_{\rm{W}}$ is positive and negative at lower and higher values of $n$, respectively, at the ferrimagnetic condition. The 2.8 eV band of $\beta$-cage clusters grows at $n \gtrapprox 7.5$ in Fig.~\ref{fig:Ref-1.5}.  The increase in the  Curie constant  at $n \approx 8.5$ is roughly estimated to be $\approx\,$$1 \times 10^{-4}$ K$\,$emu/cm$^3$ which corresponds to localized magnetic moments with spin 1/2 distributed at $\approx\,$30\% of $\beta$-cages and the saturation magnetization of $\approx\,$1.5 G.  The background Curie constant $\approx\,$$3\times 10^{-4}$ K$\,$emu/cm$^3$  at $n \approx 8.5$ corresponds to $p_{{\rm{eff}}} \approx 1.7$. This value corresponds to the saturation magnetization of $\approx\,$8 G.  As explained above in K$_n$/Na$_4$K$_{8}$-LSX, the spontaneous magnetization of supercage clusters will be much smaller than $\approx\,$8 G.  At very high magnetic fields, the saturation of total magnetization is observed at 4.2 G, as shown later in Fig.~\ref{fig:MH-highH}.

The magnetization process up to high magnetic fields at 1.3 K is shown for K$_n$/Na$_{x}$K$_{12-x}$-LSX in Fig.~\ref{fig:MH-highH}, where the respective values of ($x$, $n$) are (4, 7.7), (1.5, 8.75) and (0, 8.9).  The corresponding magnetic moment per supercage (or $\beta$-cage) is indicated in the axis on the right in units of $\mu_B$. The magnetization process in K$_{7.7}$/Na$_{4}$K$_{8}$-LSX displays a weak hump around $3.5 \times 10^4$ Oe, and the saturation at 2.7 G after the clear bend at $22.7 \times 10^4$ Oe. A hump in K$_{8.75}$/Na$_{1.5}$K$_{10.5}$-LSX is unclear, but is expected around $\approx\,$$8 \times 10^4$ Oe.  The magnetization process in K$_{8.9}$/K$_{12}$-LSX displays a hump around $16 \times 10^4$ Oe, and the saturation at $\approx\,$6 G after the bend at $\approx\,$$32 \times 10^4$ Oe.  As discussed later in Section~\ref{sec:Ferrimagnetism}, the magnetization process of ferrimagnetism in the model of classical magnetic moment has a flat magnetization up to the spin-flop field, and a constant slope up to the saturation field. The observed results, however, have round shapes at the beginning of magnetization and above the spin-flop field. This shape is explained by the increase in magnetization of the itinerant electron ferromagnetism of the supercage clusters.

%%%%%%%%%%%%%%%%%%%%%%%%%%%%%%%%%%%
\subsection{\label{sec:Electrical}Electrical properties}

An electrical conductivity and its temperature dependence give an important information on carriers in solids, especially in correlated polaron systems. The electrical conductivity $\sigma$ with different types of carriers are given by 
\begin{eqnarray}
\sigma = \sum\limits_j {e{\mu _j}{N_j}},
\label{eq:conductivity}
\end{eqnarray}
where $e$, $\mu _j$ and $N_j$ are the elementary electric charge, the $j$-th carrier mobility, and the number density of $j$-th carriers, respectively.  There are following two limiting models in the electrical conductivity having the Arrhenius law \cite{Ziese1998}.  In the band gap model with nearly temperature independent mobility, the conductivity is proportional to the number density of thermally activated free carriers $N_j$, and is expressed by the Arrhenius law. The gap energy is given by two times of the thermal activation energy. In the small polaron hopping model, the Arrhenius law can be applied to the temperature dependence of mobility approximately, where the thermal activation energy is related to the polaron formation energy, etc. A disorder and an electron correlation can have important contributions to the electrical conductivity in addition to above mentioned mechanisms.  The electrical resistivity $\rho$ is given by $1/\sigma$.

\begin{figure}[ht]
\includegraphics[width=7.0cm]{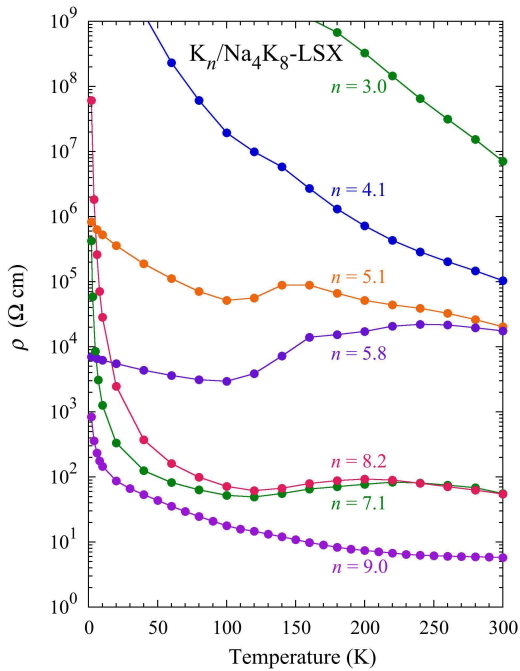}
\caption{\label{fig:rho-x=4}(Color online) Temperature dependences of the electrical resistivity $\rho$ in K$_n$/Na$_4$K$_{8}$-LSX at various values of $n$, where temperatures are decreased from 300 K.  The value of $n$ is indicated for each curve.}
\end{figure}

\begin{figure}[ht]
\includegraphics[width=6.5cm]{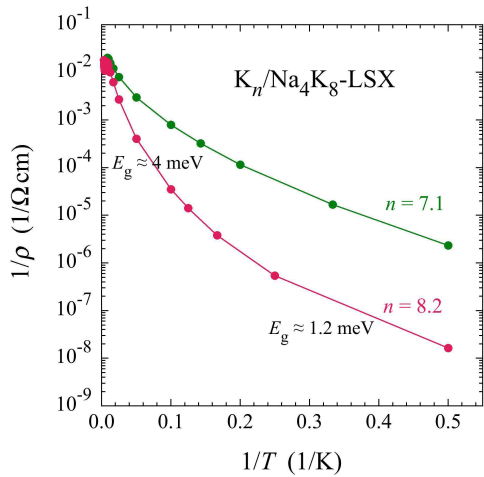}
\caption{\label{fig:sigma-n_x=4}(Color online) Temperature dependences of electrical conductivity $1/\rho$ at $n = 7.1$ and 8.2 in K$_n$/Na$_4$K$_{8}$-LSX. }
\end{figure}

\begin{figure}[ht]
\includegraphics[width=6.5cm]{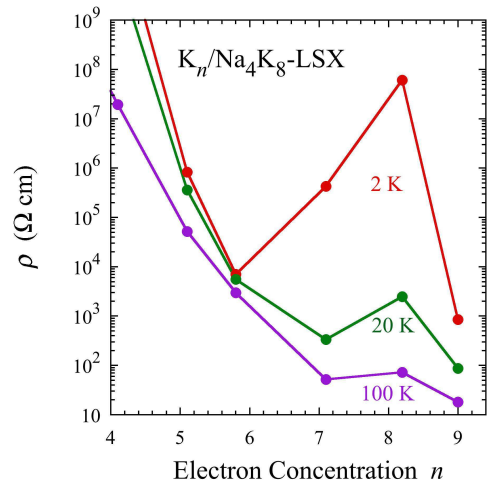}
\caption{\label{fig:rho-n_x=4}(Color online) $n$-dependence of electrical resistivity $\rho$ at 2, 20 and 100 K in K$_n$/Na$_4$K$_{8}$-LSX. }
\end{figure}

The temperature dependences of $\rho$ in K$_n$/Na$_4$K$_{8}$-LSX at various values of $n$ are shown in Fig.~\ref{fig:rho-x=4}.  The value of $n$ is indicated for each curve. The temperature of sample was decreased from 300 K. The value of $\rho$ at 300 K decreases with $n$. With the decrease in temperature, $\rho$ basically increases, because of the decrease in the mobility of small polaron hopping.  A weak anomaly is seen around 150 K.  A similar anomaly and a temperature hysteresis in $\rho$ have been clearly observed around 150 K in K$_n$/Na$_{7.3}$K$_{4.7}$-LSX \cite{Kien2015}.   In Fig.~\ref{fig:rho-x=4}, $\rho$ at $n = 5.1$ and 5.8 slightly increases at low temperatures, and is finite at the lowest temperature 2 K.  This result indicates that a finite number of free carriers (large polarons) are distributed at low temperatures.  Although $\rho$ at $n =7.1$ and 7.8 is much lower than that at $n = 5.1$ or 5.8 above $\approx\,$50 K, $\rho$ at $n =7.1$ and 7.8 quickly increases at very low temperatures and exceeds values at $n = 5.1$ or 5.8.  The electrical conductivity $\sigma = 1/\rho$ in K$_n$/Na$_4$K$_{8}$-LSX is plotted for $n = 7.1$ and 8.2 in Fig.~\ref{fig:sigma-n_x=4} as a function of the reciprocal of temperature, $1/T$. The thermal activation energy depends on temperature. The activation energy $E_{\rm{g}}$ is roughly estimated to be $\approx$1.2 and $\approx$4 meV around 3 and 15 K, respectively, for $n = 8.2$.

The $n$-dependence of $\rho$ in K$_n$/Na$_4$K$_{8}$-LSX is plotted for 2, 20 and 100 K in Fig.~\ref{fig:rho-n_x=4}.  The value of $\rho$ at 2 K decreases with $n$ up to $n = 5.8$, but increases extremely at $n =7.1$ and 8.2 at the ferrimagnetic condition $6.5 < n < 8.5$ shown in Figs.~\ref{fig:MT_x=4} and \ref{fig:x=4TcC}.  The value of $\rho$ at 2K for $n = 8.2$ is $\approx\,$$10^6$ times of that at 100 K. The increase is not significant at $n = 9.0$.  A similar increase in $\rho$ at low temperatures has been observed in K$_n$/K$_{12}$-LSX at the ferrimagnetic condition of $n$, and the value of $\rho$ at 2K for $n = 9.0$ is $\approx\,$$10^2$ times of that at 100 K \cite{Nakano2017-APX, Nakano2013K-LSX}.

%%%%%%%%%%%%%%%%%%%%%%%%%%%%%%%%%%%
\section{Discussions}

\subsection{\label{sec:tUSn}Model of correlated polaron system}

If $s$-electron wave functions of alkali metal clusters are well localized quantum-mechanically in zeolite cages, the tight-binding approximation can be applied to them \cite{Arita2004, Aoki2004}.  A narrow energy band of $s$-electrons with a strong electron correlation is expected in the supercage clusters in K$_n$/K$_{12}$-LSX, because of a large mutual Coulomb repulsion energy within supercages and the electron transfer through 12R windows \cite{Araki2019}.  Furthermore, $s$-electrons have an interaction with the displacement of alkali cations distributed in cages. Hence, $s$-electrons have an electron-phonon deformation-potential interaction as well as the electron correlation.  

In order to take an overview of the electronic properties of alkali metals in zeolites, it is effective to introduce following coarse-grained parameters of the correlated polaron system given by the so-called Holstein-Hubbard Hamiltonian \cite{Nakano2017-APX, Anderson1975, Shinozuka1987}
\begin{eqnarray} 
\begin{aligned}
H &= -\sum\limits_{i,j,\sigma } {{t_{ij}}a_{i\sigma }^\dag {a_{j\sigma }}}  + U\sum\limits_i {{n_{i \uparrow }}{n_{i \downarrow }}} \\
&+ \sum\limits_i {\left( {\frac{{P_i^2}}{{2m}} + \frac{1}{2}m{\omega ^2}Q_i^2} \right)}  - \lambda \sum\limits_i {{Q_i}\left( {{n_{i \uparrow }} + {n_{i \downarrow }}} \right),}
\end{aligned}
\label{HHH}
\end{eqnarray}
where $a_{i\sigma }$ ($a_{i\sigma }^\dag$)  is the annihilation (creation) operator of the electron with the spin $\sigma$ at the $i$-th site, and ${n_{i\sigma }} = a_{i\sigma }^\dag {a_{i\sigma }}$. $t_{ij}$ is the electron transfer energy between the $i$-th and the $j$-th sites.  $U$ is the on-site Coulomb repulsion energy (the Hubbard $U$).  The localized phonons (Einstein phonons) with the mass $m$ and the frequency $\omega$ are assumed in the third term.  $Q_i$ and $P_i$ are the lattice distortion and the conjugated momentum at the $i$-th site, respectively.  In the last term, the on-site electron-phonon interaction is introduced by the assumption of the site diagonal coupling constant $\lambda$. Here, we define the lattice relaxation energy $S$ as \cite{Shinozuka1987}
\begin{eqnarray} 
S = \frac{{{\lambda ^2}}}{{m{\omega ^2}}}.
\end{eqnarray}
If we consider the electron transfer between the nearest neighbor sites for $\left\langle {i,j} \right\rangle $ only, the first term of the right-hand side of Eq.~(\ref{HHH}) can be written as
\begin{eqnarray} 
- t\sum\limits_{\left\langle {i,j} \right\rangle ,\sigma } {a_{i\sigma }^\dag {a_{j\sigma }}},
\end{eqnarray}
where $t$ $(> 0)$ is the transfer energy of electron to the nearest neighbor site. The $t$-$U$-$S$-$n$ coarse-grained model of correlated polaron system is introduced to alkali-metal loaded zeolites, where $n$ is the average number of electrons provided by alkali atoms per site.  

\begin{figure}[ht]
\includegraphics[width=8cm]{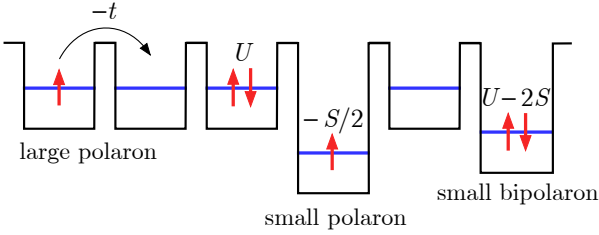}
\caption{\label{fig:HubbardModel}(Color online) Schematic illustration of the Holstein-Hubbard model.  Red arrows indicate spins of electrons. If $t$ is large enough, large polarons are stabilized as free carriers. In cases of $U > S$ and $U < S$ at small $t$, small polaron with the energy $-S/2$ and small bipolaron with the energy $U-2S$, respectively, are stable.}
\end{figure}

Schematic illustration of the Holstein-Hubbard model is given in Fig.~\ref{fig:HubbardModel}.  Red arrows indicate spins of electrons. If $t$ is large enough, large polarons migrate as free carriers.    In cases of $U > S$ and $U < S$ at small $t$, small polaron with the energy $-S/2$ and small bipolaron with the energy $U-2S$, respectively, become stable as the self-trapped states.  Small polarons and small bipolarons contribute to the conductivity by their hopping process at finite temperatures.

In zeolites, $t$ is introduced through windows between adjoining cages.  The energy band width $2B$ is given by $2B = 2ht$, where $h$ is the number of nearest neighbor sites.  $h$ is 4 for supercage or $\beta$-cage in zeolite LSX.  The energy of the band bottom is located at $-B$.  If $B < S/2$, an electron relaxes into small polaron. The value of $2B$ for supercage network is roughly estimated to be $\approx$2 eV for 1$p$ states in LSX from the spectral width of the supercage band in Fig.~\ref{fig:Abs}. $t$ for $\beta$-cage network is negligibly small, because of the large separation by D6Rs as shown in Fig.~\ref{fig:LSX-poly}.  Therefore, clusters generated in $\beta$-cages relax into self-trapped states because of a finite $S$, and become small polarons with magnetic moment or small bipolarons without magnetic moment, as discussed in Section~\ref{sec:Cluster-beta}.  

Two $s$-electrons in the same cage have a Coulomb repulsion energy $U$.  The value of $U$ depends on the size of cage, but is almost independent of the configuration of cations. The unscreened $U$ between two electrons in the 1$s$ state is estimated to be $\approx\,$3 eV for supercage with the inside diameter of $\approx\,$13 \AA{}  and $\approx\,$6 eV for $\beta$-cage with the inside diameter of $\approx\,$7 \AA{} \cite{Nakano2017-APX}.  A finite screening effect reduces the value of unscreened $U$.  A qualitative interpretation has been given by the $t$-$U$-$S$-$n$ model for various properties of alkali metals in different zeolites \cite{Nakano2017-APX}.  

At lower loading densities, $t$ is relatively small because $s$-electrons occupy lower quantum states of clusters, such as 1$s$ states, and the electron-phonon interaction $S$ dominates the system.  Hence, small bipolarons are stabilized at lower loading densities. A gap energy $\approx\,$0.6 eV in absorption spectra of dilutely K-loaded K$_{n}$/Na$_{x}$K$_{12-x}$-LSX in Fig.~\ref{fig:Abs} is assigned to the formation energy of small bipolarons at 1$s$ states in supercages.  An effective value of $t$ for the energy band near the Fermi energy is expected to increase with $n$, because $s$-electrons occupy higher quantum states of clusters, such as 1$p$ and 1$d$ states, and the metallic states are realized at large $n$ depending on the kind of alkali metals, etc. \cite{Nakano2017-APX}.  A metallic state is expected at $n \gtrapprox 6$ in K$_n$/Na$_4$K$_{8}$-LSX as shown in Fig.~\ref{fig:rho-x=4}, indicating that free carriers of large polarons are generated by 1$p$ electrons in supercage clusters.  A similar metallic transition has been observed in K$_n$/K$_{12}$-LSX  \cite{Nakano2017-APX, Nakano2013K-LSX}.

%%%%%%%%%%%%%%%%%%%%%%%%%%%%%%%%%%%
\subsection{\label{sec:Cluster-beta}Clusters at $\beta$-cages}

\begin{figure}[ht]
\includegraphics[width=8.5cm]{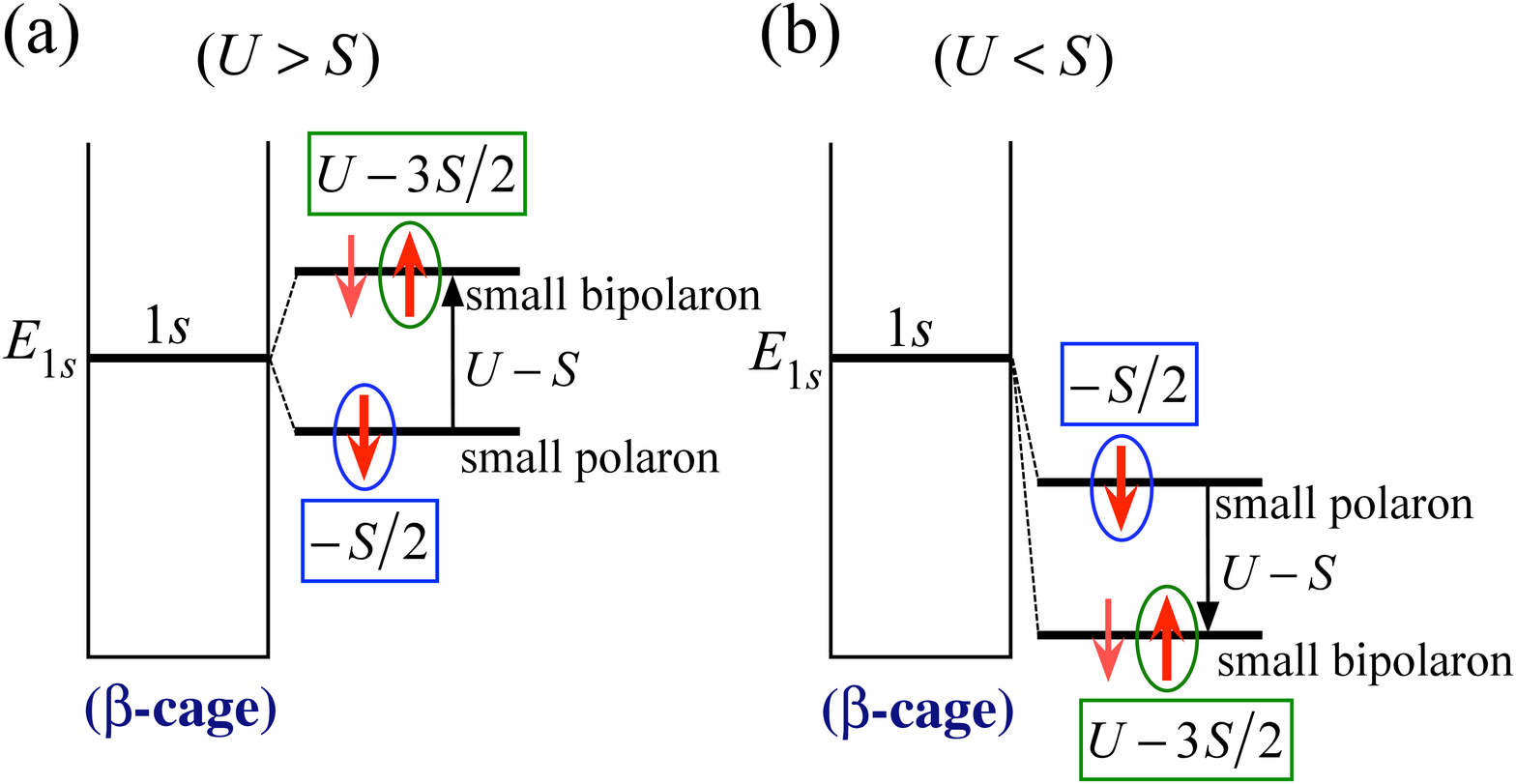}
\caption{\label{fig:bipolaron}(Color online) Schematic illustration of electronic configurations of clusters in $\beta$-cages at (a) $U > S$ and (b) $U < S$, according to the Holstein-Hubbard model.  Red arrows indicate spins of electrons. The value of $t$ is negligibly small between $\beta$-cages. Small polarons and small bipolarons in $\beta$-cages are formed depending on the relative magnitudes of $U$ and $S$ and the Fermi energy $E_{\rm{F}}$. See text in detail.}
\end{figure}

Magnetic moments of clusters in $\beta$-cages play a crucial role in magnetisms of K$_n$/Na$_x$K$_{12-x}$-LSX.  The value of $t$ between $\beta$-cages is negligibly small.  If an electron occupies the 1$s$ empty state with the energy $E_{1s}$ at $\beta$-cage, a small polaron with the energy $E_{1s} - S/2$ is generated by the electron-phonon interaction according to the Holstein-Hubbard model, as illustrated in Fig.~\ref{fig:bipolaron}.  If the second electron occupies the small polaron site, the second electron has the energy $E_{1s} + U - 3S/2$.  As shown in Fig.~\ref{fig:bipolaron}(a), small polarons with magnetic moments are generated in $\beta$-cages at $U > S$, if the Fermi energy $E_{\rm{F}}$ satisfies  
\begin{eqnarray} 
E_{1s} - \frac{S}{2} < E_{\rm{F}} < E_{1s} + U - \frac{3S}{2}.
\label{polaron}
\end{eqnarray}
With the increase in $E_{\rm{F}}$ with $n$, small bipolaron with the energy $2E_{1s} + U - 2S$ in the spin-singlet state is generated by the occupation of the second electron, if $E_{\rm{F}}$ satisfies 
\begin{eqnarray} 
E_{1s} + U - \frac{3S}{2}  < E_{\rm{F}}.
\label{polaron2}
\end{eqnarray}
This model means that small polarons with the magnetic moments are stabilized only at the condition given by Eq.~(\ref{polaron}) for $E_{\rm{F}}$.  On the other hand, there is no choice for $E_{\rm{F}}$ at $U < S$ in Eq.~(\ref{polaron}), and small polarons are unstable at any value of $E_{\rm{F}}$.  This is because the pairing of small polarons forms small bipolarons with the energy $2E_{1s} + U - 2S$ which is more stable than the separate pair of small polarons with the total energy $2E_{1s} - S$, as illustrated in Fig.~\ref{fig:bipolaron}(b), indicating that small polarons with the magnetic moments are not stabilized at any value of $n$ at $U < S$.  
  
The value of $S$ strongly depends on the kind of cations and their arrangement such as the number and the locations of cations.  Generally, $S$ for Na-rich cluster is larger than that for K-rich one, because of the larger ionization of Na atom.  The value of $S$ increases with the number of cations which contribute to the formation of cluster. 

\begin{figure}[ht]
\includegraphics[width=5.5cm]{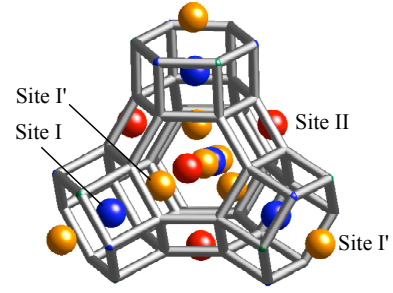}
\caption{\label{fig:site}(Color online) Schematic illustration of cation sites I, I' and II around $\beta$-cage.}
\end{figure}

Generally, cations in zeolites are located near the aluminosilicate framework, because of the attractive Coulomb force between cations and negatively charged framework. However, cations keep the mutual distance, because of the repulsive Coulomb force among them.  In each $\beta$-cage of zeolite LSX, there are three cation sites, I, I' and II, which are located at the center of D6R, the just side of D6R in $\beta$-cage and the center of 6R in supercage, respectively, as illustrated in Fig.~\ref{fig:site} \cite{Ikeda2014-Na-LSX}. There are 12 cation sites for $\beta$-cage (four sites of I, four sites of I' and four sites of II). Because site I is shared with adjoining $\beta$-cages, there are 10 cation sites per $\beta$-cage. By the loading of guest alkali metal, the number of cation increase. At the same time, the locations of cations are adjusted by the interaction with the $s$-electrons shared in clusters, as expressed by the electron-phonon interaction in the Holstein-Hubbard model.

According to the structure analysis in hydrated Na$_x$K$_{12-x}$-LSX, Na cations occupy preferably site I \cite{Lee1998}.  Sites I and II have the full occupancy, but site I' has a half occupancy.  According to the structure simulation of dehydrated zeolite LSX, the simultaneous occupations at sites I and I' are expected, unlikely in other zeolites \cite{Gibbs2002, Guesmi2012}.  The total average number of cations is $\approx\,$10 for one $\beta$-cage, and the average number becomes $\approx\,$8 per $\beta$-cage because of the sharing of site I between adjoining $\beta$-cages.  The 8 of 12 cations are distributed around each $\beta$-cage. Other 4 cations are distributed in supercage.   

By the loading of guest alkali metal, $s$-electrons are shared with cations, and the metallic bonding among cations stabilizes cation-rich clusters in $\beta$-cages.   In Na$_n$/Na$_{12}$-LSX, a full occupation of Na cations are observed simultaneously at sites I, I' and II for $n = 9.4$ and 16.7, namely 12 cations for each $\beta$-cage cluster (10 cations per $\beta$-cage cluster) \cite{Ikeda2014-Na-LSX}. In the simplest model, the possible numbers of cations for the cluster in $\beta$-cage are 10, 11 and 12 with the increase in $n$, where the numbers of cations at site I' are 2, 3 and 4, respectively.  According to this model, three kinds of $\beta$-cage clusters are expected with respective optical excitation energies.  The optical excitation energy from 1$s$ to 1$p$ states is mainly determined by the confinement potential size of $s$-electrons. The size is basically determined by that of $\beta$-cage, but these additional cations can extend slightly the effective size of the confinement potential. The origin of the different excitation energies of $\beta$-cage clusters around 2.5 eV in Figs.~\ref{fig:Abs}, \ref{fig:Ref-4} and \ref{fig:Ref-1.5} is assigned to the difference in the number of cations and the kind of cations. 

In an Na-K alloy system, a stronger cohesion effect for Na atoms makes Na-rich clusters more stable \cite{Kien2015}.  Na clusters in Na$_{12}$-LSX are nonmagnetic, because of a large $S$ \cite{Nozue2012-Na-LSX}.  At $n < 6.5$ in K$_n$/Na$_4$K$_{8}$-LSX, Na-rich clusters are expected to be stabilized at $\beta$-cages as small bipolarons at the condition of  $U < S$ in Fig.~\ref{fig:bipolaron}(b). The 2.3 eV reflection band in Fig.~\ref{fig:Ref-4} are assigned to such Na-rich small bipolarons. The candidate of magnetic clusters of small polarons is K-rich ones. At $6.5 < n < 8.5$, K-rich small polarons are expected to be stabilized at the condition of Eq.~(\ref{polaron}) for $U > S$ in Fig.~\ref{fig:bipolaron}(a), and are observed at 2.8 eV reflection band in Fig.~\ref{fig:Ref-4}, in addition to Na-rich small bipolarons at 2.3 eV. At $n > 8.5$, K-rich small bipolarons are stabilized at the condition of Eq.~(\ref{polaron2}).  
 
 At higher K-loading densities by the pressure loading in K$_n$/Na$_4$K$_{8}$-LSX, a new ferrimagnetism has been observed at the loading pressure of $\approx\,$0.5 GPa \cite{Nam2010}. The Curie constant is $\approx\,$$3.5 \times 10^{-4}$ K$\,$emu/cm$^3$ which is assigned to the contribution of magnetic sublattices of $\beta$-cage clusters and supercage ones. The spontaneous magnetization is much smaller than that expected from the Curie constant, because of the cancellation of magnetizations by the antiferromagnetic interaction between two magnetic sublattices in ferrimagnetism. The magnetic moments of $\beta$-cage clusters under the pressure loading are assigned to small polarons at 1$p$ states. 
 
 In K$_n$/Na$_{7.3}$K$_{4.7}$-LSX, the increase in localized magnetic moments have been observed clearly at $8.2 < n < 9.7$ in the increase in the Curie constant, and a nearly pure ferromagnetism has been observed at $8.4 < n < 9.7$ in the insulating phase \cite{Kien2015}.  Simultaneously, a reflection band of $\beta$-cage clusters at 2.8 eV has been observed at $n > 8$.  The origin of the magnetism is assigned to the ferromagnetic superexchange coupling between magnetic moments of $\beta$-cage clusters (small polarons) through $sp^3$ closed-shell clusters in supercages.  In reflection spectra, $\beta$-cage clusters are observed at 2.4 eV for $n \gtrapprox 4$ \cite{Kien2015}.  These clusters are nonmagnetic and and assigned to the cace of $U < S$ shown in Fig.~\ref{fig:bipolaron}(b), where Na-rich clusters are preferentially stabilized.   Clusters observed at 2.8 eV at $8 < n  \lessapprox 9.7$ are assigned to K-rich ones (small polarons) with magnetic moments at $\beta$-cages for $U > S$, and they become nonmagnetic (small bipolarons) at $n \gtrapprox 9.7$, as illustrated in Fig.~\ref{fig:bipolaron}(a). 
 
In K$_n$/K$_{12}$-LSX, pure K clusters in $\beta$-cages can be magnetic (small polarons) at large $n$. A ferrimagnetism by the antiferromagnetic interaction between localized moments of $\beta$-cage clusters and the itinerant electron ferromagnetism of supercage clusters has been observed at $n \approx 9$ \cite{Nakano2017-APX, Nakano2013K-LSX}. This ferrimagnetism disappears at $n \approx 11$ by the pressure loading at $\approx\,$0.3 GPa, because of the generation of nonmagnetic $\beta$-cage clusters (small bipolarons) \cite{Araki2019}.

%%%%%%%%%%%%%%%%%%%%%%%%%%%%%%%%%%%

\subsection{\label{sec:SublatticeInteraction}Supercage clusters and their interaction with $\beta$-cage clusters}

\begin{figure}[ht]
\includegraphics[width=7.0cm]{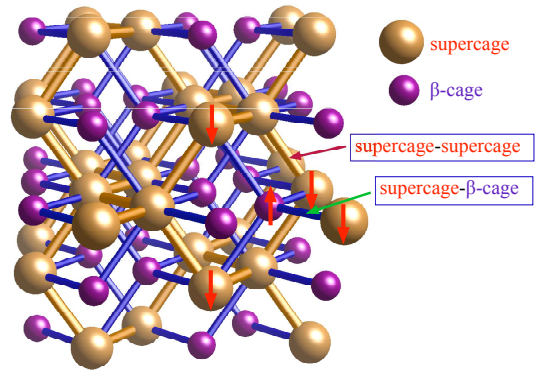}
\caption{\label{fig:bonding-b}(Color online) Schematic illustration of cluster networks in zeolite LSX. Clusters at supercages have an interaction network of a diamond structure.  Each clusters at $\beta$-cages has an interaction with clusters at four adjoining supercages. The direct interaction between $\beta$-cage clusters is absent.  See text in detail.}
\end{figure}

In zeolite sodalite (SOD framework structure), $\beta$-cages are arrayed in a body centered cubic structure by the sharing of 6Rs with eight adjoining $\beta$-cages.  An antiferromagnetism of clusters in $\beta$-cages of sodalite has been observed clearly by the antiferromagnetic interaction through 6Rs \cite{Nakano2017-APX, Srdanov1998, Sankey1998, Blake1998, Blake1999, Heinmaa2000, Madsen2001, Tou2001, Scheuermann2002, Madsen2004, Nakamura2009, Nakano2010, Nakano2012-SOD, Nakano2013JKPS, Nakano2013PRB, Nakano2015Mossbauer}.  In zeolite LSX, each $\beta$-cage shares 6Rs with four adjoining supercages.  The antiferromagnetic interaction between $\beta$-cage clusters and supercage ones occurs through 6Rs, where one up-spin in $\beta$-cage arranges down-spins in four adjoining supercages, as illustrated in Fig.~\ref{fig:bonding-b}. These four supercages with a common adjoining $\beta$-cage are the second nearest neighbors with each other. Each supercage shares 12Rs with four adjoining supercages, and electrons in supercage clusters itinerate over many supercages as large polarons. If the number density of magnetic $\beta$-cage clusters increases, the long range magnetic ordering of an itinerant electron ferromagnetism at supercage clusters is assisted geometrically by the antiferromagnetic interaction with the magnetic moments of $\beta$-cage clusters.  At the same time, the magnetic moments of $\beta$-cage clusters are ordered in the ferrimagnetism, although the direct interaction between $\beta$-cage clusters is absent. The hybridization effect of $\beta$-cage clusters with itinerant electrons of supercage clusters is expected to play an important role in electrical properties, if many $\beta$-cages are filled with small polarons with magnetic moments.

In the Kondo system, localized electron spins of magnetic atoms dilutely distributed in metal have an interaction with conduction electron spins, and an electrical resistivity gradually increases at very low temperatures.   Because of the Coulomb repulsion between localized electrons at the magnetic atom, up-spins and down-spins of conduction electrons near the Fermi energy contribute equivalently to the localized electronic state, and the Kondo singlet state is formed at very low temperatures. In the Kondo lattice system, an array of magnetic atoms provide a remarkable increase in resistivity at low temperatures, as observed in a typical Kondo insulator YbB$_{12}$ \cite{Iga1998, Kasaya1985}. The resistivity decreases under high magnetic fields up to $\approx\,$$50 \times 10^4$ Oe in YbB$_{12}$ \cite{Sugiyama1988}. 

\begin{figure}[ht]
\includegraphics[width=8cm]{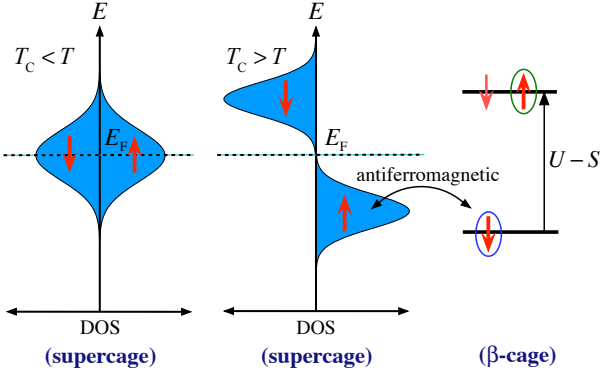}
\caption{\label{fig:DOS-ferrimag}(Color online) Schematic illustration of the energy gap model of electronic states at  ferrimagnetism ($T_{\rm{C}} > T$) and paramagnetism ($T_{\rm{C}} < T$) in K$_n$/Na$_x$K$_{12-x}$-LSX. Localized electrons in $\beta$-cages have an antiferromagnetic interaction with itinerant electrons of narrow energy band of supercage clusters. The gap is opened at the Fermi energy $E_{\rm{F}}$ at the ferrimagnetism.}
\end{figure}

In K$_n$/Na$_4$K$_{8}$-LSX, a remarkable increase in resistivity is observed at low temperatures in Fig.~\ref{fig:rho-x=4}.  This result resembles the Kondo insulator YbB$_{12}$.  A similar increase has been observed in K$_n$/K$_{12}$-LSX \cite{Nakano2017-APX, Nakano2013K-LSX}.  The activation energy indicated in Fig.~\ref{fig:sigma-n_x=4} is temperature dependent as observed in YbB$_{12}$ \cite{Iga1998, Kasaya1985}.  However, there is an essential difference between the Kondo insulator and K$_n$/Na$_x$K$_{12-x}$-LSX in magnetism. The metallic narrow band at supercage clusters in K$_n$/Na$_x$K$_{12-x}$-LSX is ferromagnetic at low temperatures both by the intraband electron-electron interaction and by the antiferromagnetic interaction with magnetic clusters at $\beta$-cages. A energy gap model of the ferrimagnetism is schematically illustrated in Fig.~\ref{fig:DOS-ferrimag}.   Electrons in magnetic clusters at $\beta$-cages have an antiferromagnetic interaction with itinerant electrons at supercages, and the energy gap opens at the Fermi energy $E_{\rm{F}}$ at low temperatures.    A change of resistivity, however, is not observed under magnetic fields up to $13 \times 10^4$ Oe in K$_n$/Na$_4$K$_{8}$-LSX within the experimental accuracy, indicating that the gap in the itinerant electron ferromagnetism seems to be kept under these magnetic fields.  A detailed theory is needed to explain these results in the future.

 In Figs.~\ref{fig:x=4TcC} and \ref{fig:x=1.5TcC},  the Weiss temperature at the ferrimagnetic region is positive and negative at lower and higher values of $n$, respectively. The Weiss temperature $T_{\rm{W}}$ in the mean field theory of localized moments is given by Eq.~(\ref{f18}) in Appendix~\ref{ferri}, where the intra-sublattice mean field coefficient of $\beta$-cage clusters, $\lambda _{\beta \beta }$, is assumed to be zero. The asymmetry of $T_{\rm{W}}$ can not be explained by the $n$-dependence of the Curie constant of $\beta$-cage clusters, $C_{\beta}$, which is defined by Eq.~(\ref{f11}), because the number density of magnetic clusters in $\beta$-cages, $N_{\beta}$, is symmetric for ferrimagnetism according to the model illustrated in Fig.~\ref{fig:bipolaron}(a).    According to Eq.~(\ref{f20}), a negative value of the Weiss temperature is expected at the condition ${C_{\rm{s}}}{\lambda _{{\rm{ss}}}} < 2{C_\beta }{\lambda _{{\rm{s}}\beta }}$, where $\lambda _{{\rm{ss}}}$ and $\lambda _{{\rm{s}}\beta }$ are the intra-sublattice mean field coefficient of supercage clusters and the inter-sublattice mean field coefficient between supercage clusters and $\beta$-cage ones, respectively. $C_{\rm{s}}$ here is the Curie constant of supercage clusters in the localized moment model and is given by Eq.~(\ref{f10}). The main reason of the asymmetry of $T_{\rm{W}}$ is expected to be the increase in $\lambda _{{\rm{s}}\beta }$ with $n$.  According to the model at $U > S$ illustrated in Fig.~\ref{fig:bipolaron}(a), $\lambda _{{\rm{s}}\beta }$ increases with $n$, because the Fermi energy $E_{\rm{F}}$ increases with $n$ and then the hybridization between electrons of supercage clusters and the localized electrons at $\beta$-cage clusters increases with $n$.

%%%%%%%%%%%%%%%%%%%%%%%%%%%%%%%%%%%

\subsection{\label{sec:Ferrimagnetism}Magnetization process of ferrimagnetism}

\begin{figure}[ht]
\includegraphics[width=5.5cm]{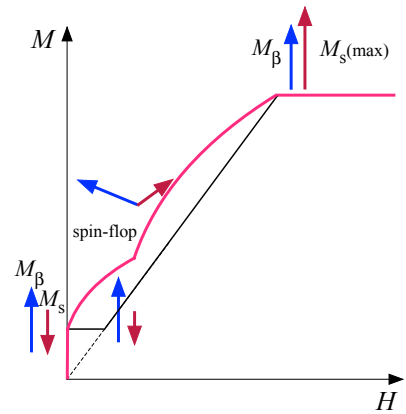}
\caption{\label{fig:mag-process}(Color online) Schematic illustration of the magnetization process of ferrimagnetism up to high magnetic fields. $M_{\beta}$ and $M_{\rm{s}}$ are magnetizations of magnetic sublattices at $\beta$-cages and supercages, respectively. See text in detail.}
\end{figure}

The magnetization process at 1.3 K in K$_n$/Na$_{x}$K$_{12-x}$-LSX shown in Fig.~\ref{fig:MH-highH} displays curves rounded out.  The magnetization process of ferrimagnetism at $T = 0$ is illustrated schematically in Fig.~\ref{fig:mag-process}.  In an ordinary ferrimagnetism of classical magnetic moments, a constant magnetization is observed up to a spin-flop field, and a constant increase in magnetization up to the saturation field, as indicated by black lines. In the ferrimagnetism in K$_n$/Na$_x$K$_{12-x}$-LSX, the magnetic sublattice at supercages is an itinerant electron ferromagnetism, and the magnetization, $M_{\rm{s}}$, increases with the applied magnetic field, because of the suppression of magnetization by the dynamical spin fluctuation \cite{Araki2019, Moriya1979, Takahashi1985, Takahashi1986, Takahashi:2017cc}.  At low fields, the dominant magnetization of the magnetic sublattice is oriented to the applied magnetic field.  For example, the dominant magnetization in the N\'eel's N-type ferrimagnetism is the magnetic sublattice at $\beta$-cages, $M_{\beta}$, below the compensation temperature. According to the mean field theory, the effective field from $M_{\beta}$ to $M_{\rm{s}}$ is opposite to the external field, and the total magnetization is given by $M_{\beta} - M_{\rm{s}}$.  With the increase in  the external field, $M_{\rm{s}}$ decreases and the total magnetization increases.  Above the spin-flop field, the angle between $M_{\rm{s}}$ and $M_{\beta}$ decreases and $M_{\rm{s}}$ increases with the external field. The total magnetization increases up to the saturation value $M_{\beta} + M_{\rm{s}}\rm{(max)}$, as indicated by red curves in Fig.~\ref{fig:mag-process}.  In K$_{7.7}$/Na$_{4}$K$_{8}$-LSX, $M_{\beta} - M_{\rm{s}} \approx 0.3$ G at low fields, and $M_{\beta} + M_{\rm{s}}\rm{(max)} \approx 2.7$ G in Fig.~\ref{fig:MH-highH}. If we assign $M_{\beta} \approx 0.7$ G in the sudden increase in the Curie constant in Fig.~\ref{fig:x=4TcC}, $M_{\rm{s}}$ at low fields and $M_{\rm{s}}\rm{(max)}$ are estimated to be $\approx$0.4 and $\approx$2.0 G, respectively. 

%%%%%%%%%%%%%%%%%%%%%%%%%%
\section{Summary}

We measured electronic properties in detail for K$_{n}$/Na$_{x}$K$_{12-x}$-LSX mainly for $x = 4$.  Ferrimagnetic properties are observed in K$_n$/Na$_4$K$_{8}$-LSX and  K$_n$/Na$_{1.5}$K$_{10.5}$-LSX. At the same time, the Curie constant increases, and a reflection band of $\beta$-cage clusters at 2.8 eV is observed in accordance with the ferrimagnetism.  An electrical resistivity indicates metallic value at  $n \gtrapprox 6$ in K$_n$/Na$_4$K$_{8}$-LSX. The ferrimagnetism is explained by the antiferromagnetic interaction between the magnetic sublattice of localized moments at $\beta$-cage clusters and that of itinerant electron ferromagnetism at supercage clusters.  The electrical resistivity increases extraordinarily at low temperatures in ferrimagnetic samples.  We try to explain the anomaly in the electrical resistivity by the analogy of the Kondo insulator, where itinerant electrons of supercage clusters interact with localized electrons of $\beta$-cage clusters.  However, itinerant electrons of the narrow energy band of supercage clusters are ferromagnetic, differently from nonmagnetic electrons of the ordinary energy band in the Kondo insulator.

%%%%%%%%%%%%%%%
\begin{acknowledgments}
We are deeply grateful to Profs. R. Arita, K. Nakamura, and H. Aoki for theoretical studies and discussions.   We also thank Mr. S. Tamiya (Osaka University) for chemical analysis.  This work was supported by Grant-in-Aid for Scientific Research on Priority Areas (No. JP19051009), Grant-in-Aid for Scientific Research (A) (No. JP24244059 and No. JP13304027) and (C) (No. JP26400334), Grant-in-Aid for Creative Scientific Research ``New Phases of Matter in Multidisciplinary Approaches" (No. JP15GS0213), Global COE Program ``Core Research and Engineering of Advanced Materials-Interdisciplinary Education Center for Materials Science'' (G10), the 21st Century COE Program ``Towards a new basic science: depth and synthesis'' (G17), MEXT Japan.

\end{acknowledgments}

\appendix

\section{Ferrimagnetism}
\label{ferri}

We calculate a ferrimagnetism by the use of the mean field (molecular field) theory. We assume two nonequivalent magnetic sublattices of localized moments corresponding to supercage clusters and $\beta$-cage ones in K$_{n}$/Na$_{x}$K$_{12-x}$-LSX.  The geometrical arrangement shown in Fig.~\ref{fig:bonding-b} and the itinerant electron ferromagnetism of supercage clusters are not considered.   

We define mean fields for supercage clusters and $\beta$-cage clusters, $H_{{\rm{ms}}}$ and $H_{{\rm{m}}\beta }$, respectively, as 
\begin{eqnarray} 
{H_{{\rm{ms}}}} = {\lambda _{{\rm{ss}}}}{M_{\rm{s}}} - {\lambda _{{\rm{s}}\beta }}{M_\beta },
\label{f1}
\\
{H_{{\rm{m}}\beta }} = {\lambda _{\beta \beta }}{M_\beta } - {\lambda _{{\rm{s}}\beta }}{M_{\rm{s}}},
\label{f2}
\end{eqnarray}
where $M_{\rm{s}}$ and $M_\beta$ are magnetizations of respective magnetic sublattices for the ferrimagnetism, and $\lambda _{{\rm{ss}}}$, $\lambda _{\beta \beta }$ and $\lambda _{{\rm{s}}\beta }$ the intra-sublattice mean field coefficient of supercage clusters, that of $\beta$-cage clusters and the inter-sublattice mean field coefficient between supercage clusters and $\beta$-cage ones, respectively.  The minus sign of the second term in the right hand side of above equations means an antiferromagnetic interaction between two magnetic sublattices. 

The magnetizations of both sublattices under the external magnetic field $H$ at the temperature $T$ are given as
\begin{eqnarray} 
{M_{\rm{s}}} = {N_{\rm{s}}}{g_{\rm{s}}}{\mu _{\rm{B}}}{B_{{J_{\rm{s}}}}}\left( {{g_{\rm{s}}}{\mu _{\rm{B}}}{J_{\rm{s}}}\frac{{H + {H_{{\rm{ms}}}}}}{{{k_{\rm{B}}}T}}} \right),
\label{f3}
\\
{M_\beta } = {N_\beta }{g_\beta }{\mu _{\rm{B}}}{B_{{J_\beta }}}\left( {{g_\beta }{\mu _{\rm{B}}}{J_\beta }\frac{{H + {H_{{\rm{m}}\beta }}}}{{{k_{\rm{B}}}T}}} \right),
\label{f4}
\end{eqnarray}
where $N_{\rm{s}}$ and $N_\beta$ are the number densities of supercage clusters and $\beta$-cage ones, respectively, $g_{\rm{s}}$ and $g_\beta$ the $g$ values of respective clusters, and $J_{\rm{s}}$ and $J_\beta$ the total angular momentum quantum numbers of respective clusters.  $k_{\rm{B}}$ is the Boltzmann constant.  $B_{J}(y)$ is the Brillouin function, and is given for $\left| y \right| \ll 1$ as
\begin{eqnarray} 
B_{J}(y) = \frac{J+1}{3J}y.
\label{f5}
\end{eqnarray}
At sufficiently high temperatures of paramagnetism, following conditions are satisfied:
\begin{eqnarray} 
\left| {{g_{\rm{s}}}{\mu _{\rm{B}}}{J_{\rm{s}}}\frac{{H + {H_{{\rm{ms}}}}}}{{{k_{\rm{B}}}T}}} \right| \ll 1,
\label{f6}
\\
\left| {{g_\beta }{\mu _{\rm{B}}}{J_\beta }\frac{{H + {H_{{\rm{m}}\beta }}}}{{{k_{\rm{B}}}T}}} \right| \ll 1.
\label{f7}
\end{eqnarray}
Then, we obtain following magnetizations by using Eqs.~(\ref{f1}) and (\ref{f2}) as
\begin{eqnarray} 
{M_{\rm{s}}} = \frac{{{C_{\rm{s}}}}}{T}\left( {H + {\lambda _{{\rm{ss}}}}{M_{\rm{s}}} - {\lambda _{{\rm{s}}\beta }}{M_\beta }} \right),
\label{f8}
\\
{M_\beta } = \frac{{{C_\beta }}}{T}\left( {H + {\lambda _{\beta \beta }}{M_\beta } - {\lambda _{{\rm{s}}\beta }}{M_{\rm{s}}}} \right),
\label{f9}
\end{eqnarray}
where the Curie constants of supercage clusters and $\beta$-cage ones, $C_{\rm{s}}$ and $C_\beta$, respectively, are given as
\begin{eqnarray} 
{C_{\rm{s}}} = \frac{{{N_{\rm{s}}}g_{\rm{s}}^2\mu _{\rm{B}}^2{J_{\rm{s}}}\left( {{J_{\rm{s}}} + 1} \right)}}{{3{k_{\rm{B}}}}},
\label{f10}
\\
{C_\beta } = \frac{{{N_\beta }g_\beta ^2\mu _{\rm{B}}^2{J_\beta }\left( {{J_\beta } + 1} \right)}}{{3{k_{\rm{B}}}}}.
\label{f11}
\end{eqnarray}

From Eqs.~(\ref{f8}) and (\ref{f9}), the total magnetic susceptibility $\chi$ is given by
\begin{eqnarray} 
\begin{aligned}
&\chi  = \frac{{{M_{\rm{s}}}}}{H} + \frac{{{M_\beta }}}{H}\\
&= \frac{{T\left( {{C_{\rm{s}}} + {C_\beta }} \right) - {C_{\rm{s}}}{C_\beta }\left( {2{\lambda _{{\rm{s}}\beta }} + {\lambda _{\beta \beta }} + {\lambda _{{\rm{ss}}}}} \right)}}{{{T^2} - T\left( {{C_{\rm{s}}}{\lambda _{{\rm{ss}}}} + {C_\beta }{\lambda _{\beta \beta }}} \right) + {C_{\rm{s}}}{C_\beta }\left( {{\lambda _{{\rm{ss}}}}{\lambda _{\beta \beta }} - {\lambda _{{\rm{s}}\beta }}^2} \right)}}.
\end{aligned}
\label{f12}
\end{eqnarray}
The Curie temperature $T_{\rm{C}}$ is obtained from Eq.~(\ref{f12}) by the divergence condition at the higher temperature as
\begin{eqnarray} 
\begin{aligned}
{T_{\rm{C}}} =& \frac{{{C_{\rm{s}}}{\lambda _{{\rm{ss}}}} + {C_\beta }{\lambda _{\beta \beta }}}}{2}\\
&+ \frac{{\sqrt {{{\left( {{C_{\rm{s}}}{\lambda _{{\rm{ss}}}} - {C_\beta }{\lambda _{\beta \beta }}} \right)}^2} + 4{C_{\rm{s}}}{C_\beta }{\lambda _{{\rm{s}}\beta }}^2} }}{2}.
\end{aligned}
\label{f13}
\end{eqnarray}
At sufficiently high temperatures, $\chi$ is expected to approache the Curie-Weiss law
\begin{eqnarray} 
\chi  \approx \frac{{{C_{\rm{s}}} + {C_\beta }}}{{T - {T_W}}},
\label{f14}
\end{eqnarray}
where $C_{\rm{s}} + C_\beta$ and $T_{\rm{W}}$ are the total Curie constant and the Weiss temperature, respectively.  We obtain the relation at sufficiently high temperatures from Eq.~(\ref{f12}) as
\begin{eqnarray} 
\begin{aligned}
\frac{{{C_{\rm{s}}} + {C_\beta }}}{\chi } \approx T +& \frac{{{C_{\rm{s}}}{C_\beta }\left( {2{\lambda _{{\rm{s}}\beta }} + {\lambda _{\beta \beta }} + {\lambda _{{\rm{ss}}}}} \right)}}{{{C_{\rm{s}}} + {C_\beta }}} \\
&- \left( {{C_{\rm{s}}}{\lambda _{{\rm{ss}}}} + {C_\beta }{\lambda _{\beta \beta }}} \right).
\end{aligned}
\label{f15}
\end{eqnarray}
Finally, we extract $T_{\rm{W}}$ from Eqs.~(\ref{f14}) and (\ref{f15}) as
\begin{eqnarray} 
\begin{aligned}
{T_{\rm{W}}} =  - &\frac{{{C_{\rm{s}}}{C_\beta }\left( {2{\lambda _{{\rm{s}}\beta }} + {\lambda _{\beta \beta }} + {\lambda _{{\rm{ss}}}}} \right)}}{{{C_{\rm{s}}} + {C_\beta }}} \\
&+  {{C_{\rm{s}}}{\lambda _{{\rm{ss}}}} + {C_\beta }{\lambda _{\beta \beta }}} 
\end{aligned}
\label{f16}
\end{eqnarray}

If we assume no intra-sublattice interaction of $\beta$-cage clusters as ${\lambda _{\beta \beta }} = 0$, we obtain $T_{\rm{C}}$ and $T_{\rm{W}}$ as
\begin{eqnarray} 
{T_{\rm{C}}} = \frac{{{C_{\rm{s}}}{\lambda _{{\rm{ss}}}}}}{2}\left( {1 + \sqrt {1 + \frac{{4{C_\beta }\lambda _{{\rm{s}}\beta }^2}}{{{C_{\rm{s}}}\lambda _{{\rm{ss}}}^2}}} } \right),
\label{f17}
\\
{T_{\rm{W}}} = \frac{{{C_{\rm{s}}}}}{{{C_{\rm{s}}} + {C_\beta }}}\left( {{C_{\rm{s}}}{\lambda _{{\rm{ss}}}} - 2{C_\beta }{\lambda _{{\rm{s}}\beta }}} \right).
\label{f18}
\end{eqnarray}
The positive and negative values of $T_{\rm{W}}$ are obtained as
\begin{eqnarray} 
{T_{\rm{W}}} > 0{\rm{~~at~~}}{C_{\rm{s}}}{\lambda _{{\rm{ss}}}} > 2{C_\beta }{\lambda _{{\rm{s}}\beta }},
\label{f19}
\\
{T_{\rm{W}}} < 0{\rm{~~at~~}}{C_{\rm{s}}}{\lambda _{{\rm{ss}}}} < 2{C_\beta }{\lambda _{{\rm{s}}\beta }}.
\label{f20}
\end{eqnarray}

% The \nocite command causes all entries in a bibliography to be printed out
% whether or not they are actually referenced in the text. This is appropriate
% for the sample file to show the different styles of references, but authors
% most likely will not want to use it.
%\nocite{*}

%\bibliography{apssamp}% Produces the bibliography via BibTeX.

%%%%%%%%%%%%%%%%%%%%%%%%%%%%%%%%%%%%%

\end{document}